\documentclass[pre,reprint,showpacs,showkeys,samsmath,amssymb,floatfix]{revtex4-1}

\usepackage{times}
\usepackage{amsmath}    
\usepackage{amsfonts}   
\usepackage{amssymb}
\usepackage{graphicx}   
\usepackage{subfigure}
\usepackage{bbm}
\usepackage{graphicx}
\usepackage{lineno}

\newcommand{\e}{{\rm e}}
\newcommand{\ra}{\rightarrow}
\newcommand{\mx}{m_{1}}
\newcommand{\my}{m_{2}}
\newcommand{\lmx}{{m}_{1}^{(d)}}
\newcommand{\lmy}{{m}_{2}^{(d)}}
\newcommand{\amx}{{m}_{1}^{(g)}}
\newcommand{\amy}{{m}_{2}^{(g)}}
\newcommand{\mxo}{m_{1}(0)}
\newcommand{\myo}{m_{2}(0)}
\newcommand{\mxe}{m_{{\rm eq};1}}
\newcommand{\mye}{m_{{\rm eq};2}}

\newcommand{\vars}{{\sigma^2}}
\newcommand{\de}{\delta}
\def\noi{\noindent}
\def\ra{\rightarrow}

\newcommand{\bi}{\begin{itemize}}
\newcommand{\ei}{\end{itemize}}

\begin{document}

\title{Kinetic Model of Mass Exchange with Dynamic Arrhenius Transition Rates}

\author{Dionissios T. Hristopulos}
 \email{dionisi@mred.tuc.gr}   
\author{Aliki D. Muradova}
 \email{aliki@mred.tuc.gr}   
 \affiliation{School of Mineral Resources Engineering, Technical University of Crete,
Chania 73100, Greece}

\date{\today}

\begin{abstract}

We study a nonlinear kinetic model of  mass exchange between interacting grains.
The  transition rates follow the Arrhenius equation
with an activation energy that depends on the grain mass.
 We show that the activation parameter can be absorbed in the initial conditions for the grain masses,
  and that the total mass is conserved.
 We obtain numerical solutions of the
coupled, nonlinear, ordinary differential equations of mass exchange for the two-grain system,
and we compare them with approximate theoretical solutions in specific neighborhoods of the phase space.
Using phase plane methods, we determine that the system exhibits  regimes of diffusive and growth-decay (reverse diffusion) kinetics.
The equilibrium states
are determined by the  mass  equipartition and separation nullcline curves.
 If the transfer rates are perturbed by white noise, numerical simulations show that the system still exhibits diffusive and
 growth-decay regimes, although the noise can reverse the sign of equilibrium mass difference.
 Finally, we present theoretical analysis and numerical simulations of a system with
 many interacting grains. Diffusive and growth-decay regimes are established as well, but the approach to
 equilibrium is  considerably slower.
 Potential applications of the mass exchange model involve coarse-graining during sintering and wealth exchange in econophysics.
\end{abstract}



\maketitle


\newpage

\section{Introduction}
Non-equilibrium processes such as grain growth and nucleation remain a topic of interest in statistical physics~\cite{Cetinel13}.
Such phenomena  are common in many engineering and physical
processes~\cite{Rubi03}. A grain is defined as a contiguous region of material with the same
crystallographic orientation which changes discontinuously at the grain boundaries.
Many technological materials, including advanced ceramics,  are
produced by means of non-equilibrium physical processes that generate
phase changes and grain growth.
Early studies of the kinetics of crystallization and other phase changes were based on the
Johnson-Mehl-Avrami-Kolmogorov equation~\cite{avrami39,avrami41}.
The process of solid-state
sintering transforms  a powder into  a monolithic
material by applying temperature and pressure~\cite{Eggers98}.
 Sintering involves diffusion and transport
of atoms as well as plastic deformations.
During the sintering process, the number of grains is progressively reduced, while the average grain radius
 increases in a process known as Ostwald ripening~\cite{Yao93,Pototsky14}.

Whereas sintering is  essentially a simple process of densification by heating, its details are complicated. Hence, modeling
 the sintering kinetics is a topic of continuing research. Recent computational approaches
 involve Direct Multiscale Modeling~\cite{Max12}
and the Discrete Element Method (DEM), which relax assumptions regarding the
particle kinematics~\cite{martin06,martin09} and generalized Monte Carlo simulations~\cite{Cetinel13}. In DEM, grain
coarsening is incorporated by transferring the overlapping volume of
neighboring spherical particles from the smaller to the larger.
Existing sintering models are continuum formulations. This is also
true of diffusion processes such as the Cahn-Hilliard equation~\cite{CH58}, which
describes phase separation (reverse diffusion), and the phase-field models used to describe solidification~\cite{Langer86}.
Recently, a self-consistent, mean-field kinetic theory was proposed to describe atomic diffusion in
non-uniform alloys~\cite{Nastar14}. This theory uses thermally activated transition rates between species and
corrects the Cahn-Hilliard model in the presence of non-uniformities.

\begin{figure}
\centering
\includegraphics[width=0.45\linewidth]{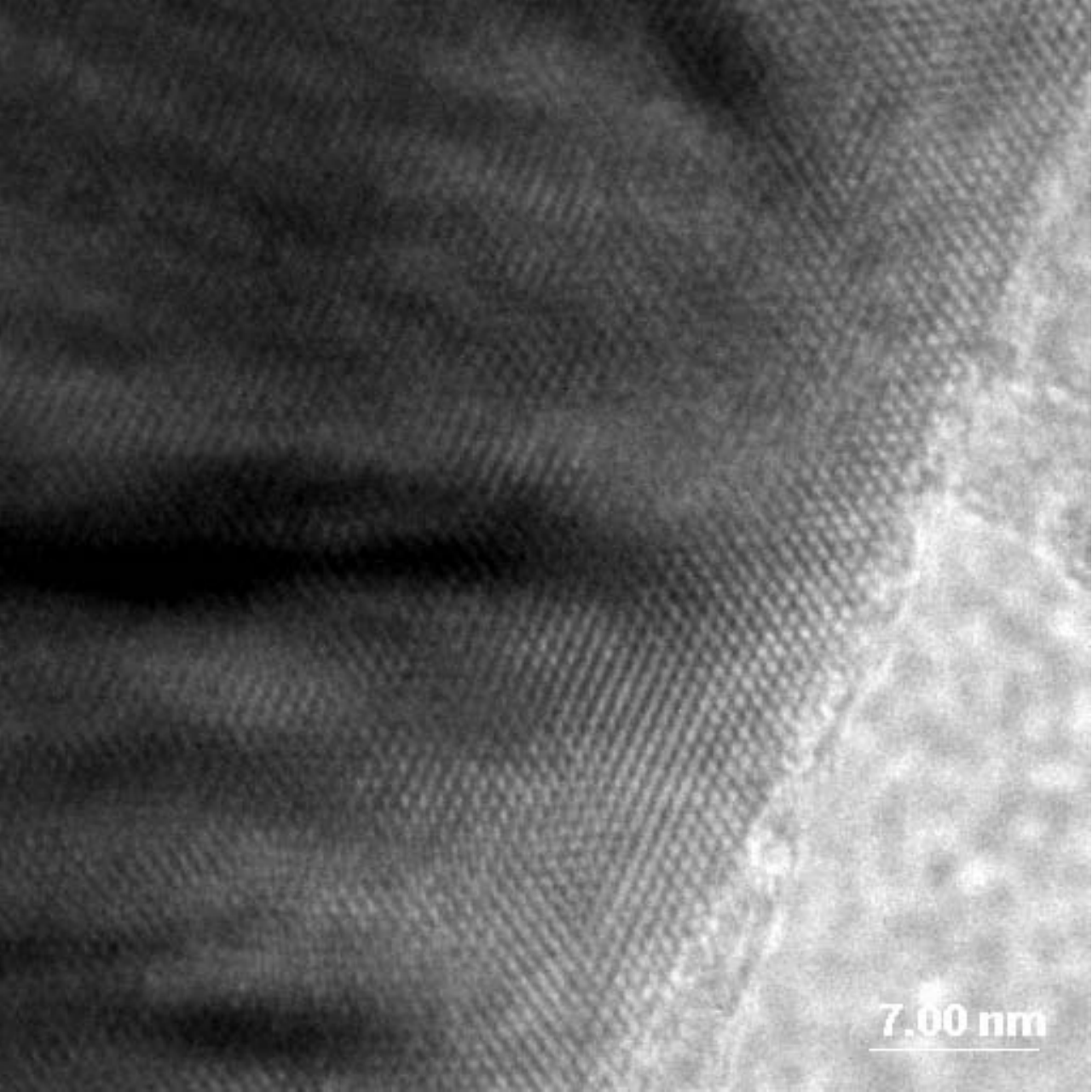}
\caption{High resolution Transmission Electron
Microscope (HRTEM) image of mechanically activated alpha-silicon nitride grain.
The image is provided courtesy of Prof. Malgorzata
Sopicka-Lizer, Silesian University of Technology,
Poland.}\label{fig:hrtem}
\end{figure}

Mass diffusion during sintering is controlled by an activation energy
which can be lowered by means of ball milling~\cite{Sopicka13}.
A high resolution transmission electron
microscope image of an alpha-silicon nitride grain
(an alloy based on  silicon nitride, Si$_{3}$N$_{4}$, in which
some silicon atoms are replaced by {Al} and corresponding  {Ni}
atoms by {O}) after mechanical activation by ball milling is shown in
Fig.~\ref{fig:hrtem}. The grain includes areas with both oriented and disordered
lattice structure which occur both inside and near the
boundary of the grain. This structure can lead to both intra-grain and inter-grain diffusion.

Motivated by the above observations, we study a simplified model of mass
exchange between  grains. The exchange is governed by a
kinetic equation which involves transition rates that are based on  the Arrhenius equation
$k = \exp(-E_{a}/k_{B}T)$, where $k$ is the rate coefficient (reaction constant),
$E_{a}$ is the activation energy, $k_{B}$ is Boltzmann's constant, and $T$
is the temperature. The kinetic model that we study herein is too simple to accurately capture
 properties of the actual sintering process.
 Nevertheless, it exhibits a notable  transition between a diffusive regime
 in which the equilibrium grain masses tend to become equal, and a growth-decay (reverse diffusion) regime in which the larger mass
 grows whereas the smaller one shrinks. These two distinct regimes may be related, respectively, to
 normal and abnormal growth regimes observed in sintering~\cite{Hillert65}.

\subsection{Nonlinear Mass Exchange Model}
\label{ssec:nl-mem}
In~\cite{dth06} we introduced a kinetic model for mass exchange
between $N$ grains of different radii.
The model involves a system of $N$ coupled,
 non-linear, ordinary differential equations (ODEs) with Arrhenius-like
transition rate coefficients. The non-dimensional form of the system is given by

\begin{align}
\label{eq:mass-trans2}
\frac{d {m}_{i}({t})}{d t}  & =
\sum_{<i,j>} \e^{  - u  \, {m}_{j}({t})   }
  {m}_{j}({t})
   -   \sum_{<i,j>}   \e^{  - u  \, {m}_{i}({t}) } {m}_{i}({t}).
\end{align}

\noindent In~\eqref{eq:mass-trans2} ${m}_{i}({t})$,
and $u$  are, respectively, the non-dimensional grain mass,
 time, and grain activation energy as defined in~\cite{dth06}, whereas the symbol $<i,j>$ denotes
that the summation is over the nearest neighbors of the $i-$th grain.
 The \emph{activation parameter} $u$ is given by
$ u = \alpha Q /N_{A} k_{B} T$, where $\alpha$ is a dimensionless rate factor,
$Q$ is the characteristic activation energy, $N_{A}$ is the Avogadro constant, $ k_{B}$ is Boltzmann's constant,
and
$T$ is the temperature. The exponent of the Arrhenius factor is also proportional
to the dimensionless mass  $m_{i}(t)$.
The activation energy is lower for grains with a higher degree
of amorphization (i.e., fraction of the grain in the amorphous state).
The presence of the grain mass in the exponent is justified by the fact that
 smaller grains  are expected
to have a higher degree of amorphization and therefore lower activation energy.

The system of eqs.~\eqref{eq:mass-trans2} focuses on
the exchange of mass between grains through transitions that
incorporate non-homogeneous activation energies but neglects plastic
deformation effects.
In~\cite{dth06}, the ODE system of eqs.~\eqref{eq:mass-trans2} was numerically solved
for a one-dimensional chain using Euler's first-order explicit method with
an adaptive step size~\cite{Press97}. A Gaussian initial distribution of grains was used, and solutions
were obtained for (i) $u=0.1$ and (ii) $u=15$. It was found that the  system
 follows two qualitatively different behaviors; in case (i)
 the masses of all the grains converge to the mean of the distribution,
 whereas in case (ii) some grain masses tend to zero and others increase.

In this work we investigate  the equilibrium states and the
dynamics of model~\eqref{eq:mass-trans2}. Section~\ref{sec:two-grain} gives an
 exploratory analysis of  the two-grain system which shows that activation parameter
 $u$ can be absorbed in the initial conditions.  Using phase-plane methods, in Section~\ref{sec:equil}
 we determine the equilibrium states of the two-grain system as a function of the initial masses. We find equipartition and
 mass separation equilibrium states, and we identify a trapping effect which impedes further evolution of the system.
Section~\ref{sec:dynamic} investigates the dynamic regimes of the model using numerical solutions
and explicit approximations. We determine a diffusive regime in which the grain masses tend to become equal and a growth-decay
regime in which the larger grain grows at the expense of the smaller, and we show that both can be interrupted
by trapping. In Section~\ref{sec:noise} we study the behavior of the two-grain system with noisy
transition rates by means of numerical simulations; we demonstrate that the noise  in the growth-decay regime  can switch the direction
of growth from the larger to the smaller grain, if
the initial masses are nearly equal. The $N$-grain system is investigated in Section~\ref{sec:N-grains}. We establish
by means of numerical solutions diffusive and growth-decay regimes with slower relaxation rates than in the two-grain system, and we find that the
grain mass evolution is not necessarily monotonic in time. We also show that the diffusion regime is obtained from the nonlinear $N$-grain system
at the limit of small initial masses.  Finally,
we present our conclusions in Section~\ref{sec:conclu}.

\section{Two-Grain Mass Exchange Model}
\label{sec:two-grain}
We investigate a two-grain system that follows eqs.~\eqref{eq:mass-trans2}
in order to understand the impact of exponentially
varying  transition rates on mass exchange. Let us consider
$m_{1}(t)= x(t)$ and $m_{2}(t) = y(t)$. Then, the mass transfer between the grains
is expressed by means of the following system of first-order, autonomous, nonlinear, ordinary differential equations (ODEs)

\begin{subequations}
\label{eq:mass-x-y}
\begin{align}
\frac{dx(t)}{dt} &  = y(t)\, \e^{-u\,y(t)} - x(t) \, \e^{-u\,x(t)}
\\
\frac{dy(t)}{dt} &  = x(t)\, \e^{-u\,x(t)} - y(t) \, \e^{-u\,y(t)},
\end{align}
\end{subequations}

\noi with initial conditions $x(0)=x_{0}$ and $y(0)=y_{0}$.

Assuming a positive \emph{activation parameter},  $u > 0$, by means of the transformations
$ u\, x(t) \mapsto m_{1}(t)$ and $ u\, y(t) \mapsto m_{2}(t)$,
eqs.~\eqref{eq:mass-x-y} transform as follows

\begin{subequations}
\label{eq:two-grain-nou}
\begin{align}
\frac{d\mx(t)}{dt} &  = f_{2,1}(t) - f_{1,2}(t)
\\
\frac{d\my(t)}{dt} &  =  f_{1,2}(t) - f_{2,1}(t),
\end{align}
\end{subequations}

\noi with initial conditions $\mx(0)= u\,x_{0}$ and  $\my(0)=u\,y_{0}$.
The  Arrhenius-like \emph{transition rates}
$f_{i, j}(t) = m_{i}(t)\, \e^{-m_{i}(t)}$ $(i,j=1,2, i\neq j)$
 are bounded from above by $1/\e$. The differences
$f_{2,1}(t) - f_{1,2}(t)$ and $f_{1,2}(t) - f_{2,1}(t)$ represent the \emph{mass transfer rates} to $\mx(t)$ and $\my(t)$, respectively.
The phase space of eqs.~\eqref{eq:two-grain-nou} is determined by the two initial
conditions which define the parameter space. Hence, parameter $u$, which is
 absorbed in the initial conditions, is irrelevant: if solutions  $\mx(t)$ and $\my(t)$ of eqs.~\eqref{eq:two-grain-nou}
with initial conditions $\mx(0)= u\,x_{0}$ and $\my(0)=u\,y_{0}$ are available, the
solution for $x(t)$ and $y(t)$ can be obtained for any $u >0$ from
$x(t) = \mx(t)/u$ and $y(t) = \my(t)/u$.

The absorption of the activation parameter
in the initial conditions allows us to focus on a two-dimensional parameter space.
It is straightforward to show that the same transformation can be applied to an $N$-grain system, leading to an
$N$-dimensional phase space.
Below we investigate the properties of system~\eqref{eq:two-grain-nou} by means of phase-plane methods which
are commonly used in the study of nonlinear ordinary differential
equations~\cite{king03}.
We use the notation $\mxe = \lim_{t \ra \infty} \mx(t)$, $\mye = \lim_{t \ra \infty} \my(t)$.

\section{Equilibrium States}
\label{sec:equil}
The ODE system~\eqref{eq:two-grain-nou} conserves the total grain
mass, as shown by adding the left- and right-hand sides of the
two equations, respectively, leading to
the cumulative mass evolution equation
$\frac{d[\mx(t)+\my(t)]}{dt} = 0.$

\begin{figure}
\includegraphics[width=0.8\linewidth]{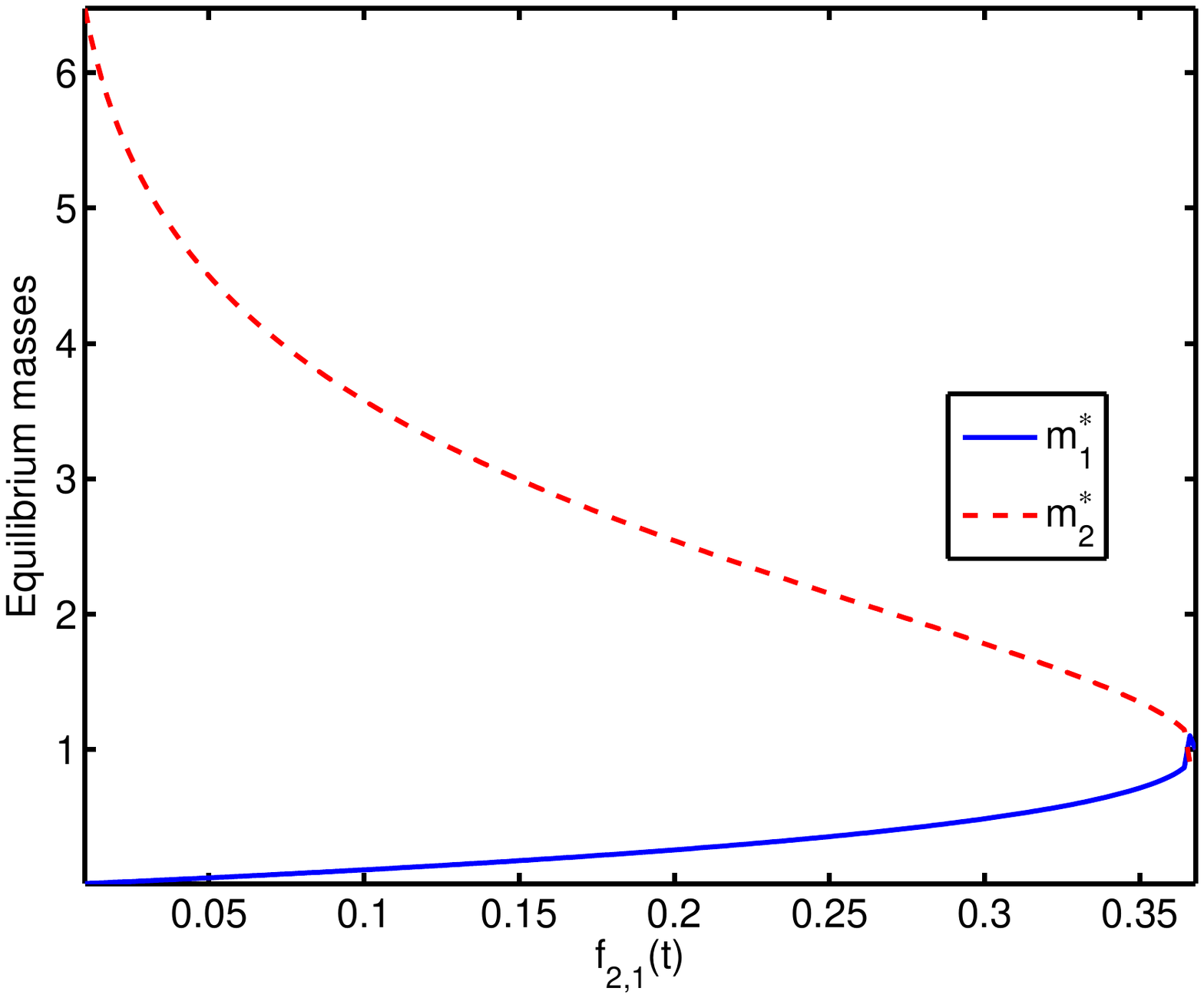}
 \caption{\label{fig:roots-rate} Roots of the equation $ m\, \e^{-m} =c$, where $ 0 \le c < 1/\e$.
 The horizontal axis shows possible values of the transition rate $f_{2,1}(t)$ ---or equivalently of
 $f_{1,2}(t)$--- at any $t$.
 The solution has two branches, implying that the
 zero-transition-rate condition is realized by two different mass values,
 $\mx^{\ast}$ and $\my^{\ast}$. The locus of these points defines the curvilinear mass separation nullcline
 shown in Fig.~\ref{fig:nullclines}.}
\end{figure}

For the ODE~\eqref{eq:two-grain-nou} the \emph{nullclines} (i.e., the curves along which the
mass derivatives  with respect to time vanish)
corresponding to $m_{1}(t)$ and
$m_{2}(t)$ are the curves defined by means of the equations $d\mx(t)/dt=0$ and $d\my(t)/dt=0$,
 respectively. It follows from the mass conservation that $d\mx(t)/dt= -d\my(t)/dt=0$, and thus the nullclines
 for the two grains coincide. Equilibrium points arise at the intersection of nullclines. Given the
 coincidence of the nullclines
 for the two equations of the system~\eqref{eq:two-grain-nou}, each nullcline consists entirely of equilibrium points.
The nullclines are given by the equation

\begin{equation*}
f_{2,1}(t)  -  f_{1,2}(t)  =0.
\end{equation*}
Let us denote the solutions of the above equation by  $m^{\ast}_{1}$ and $m^{\ast}_{2}$.
One such solution is $m^{\ast}_{1} = m^{\ast}_{2} = c$, where $c>0$.
This is the \emph{equipartition nullcline} curve.
Other equilibrium points correspond to  degenerate solutions of the nonlinear equation
 $ m\, \e^{-m} =c$, where $ 0 \le c < 1/\e$.
The dependence of the roots $m^{\ast}_{1}$ and $m^{\ast}_{2}$ of the above equation on $c$ is shown in Fig.~\ref{fig:roots-rate}:
The two roots have quite different values for $c\ll 1$, whereas they converge
as $ c \to 1$. The curvilinear trace of the points $(m^{\ast}_{1}, m^{\ast}_{2}) $ constitutes the
\emph{mass separation nullcline} which satisfies the equation
\[
m_{1} - \ln m_{1} = m_{2} - \ln m_{2}.
\]

Based on the above analysis, the equilibrium points of system~\eqref{eq:two-grain-nou} coincide with the
 nullcline curves, which constitute two \emph{equilibrium curves}.
 The nullcline $\mxe = \mye = (\mx(0) + \my(0))/2$ represents the \emph{equipartition equilibrium}.
This equilibrium state is the result of a diffusive process that redistributes
the total mass between the grains.
In contrast, the second nullcline represents a \emph{separation or reverse diffusion equilibrium}. In the latter case,
the larger grain increases its mass at the expense of the smaller grain.

\emph{Trapping} occurs if the evolution of the system is arrested at the separation nullcline.
The system is  ``frozen'' (trapped) at the outset if the initial conditions satisfy the
nullcline conditions, i.e., $\mx(0)= \my(0)$ or $\mx(0) \, \e^{-\mx(0)}= \my(0)\, \e^{-\my(0)}$.
For other initial conditions, the two-grain system evolves toward one of the
two equilibrium curves. As we show below, trapping by the separation nullcline can occur for systems that
 evolve either in the growth-decay or the diffusive regime.
For all practical purposes, trapping  also occurs if $\mx(0), \my(0) \gg 1$,
because the transition rates in eqs.~\eqref{eq:two-grain-nou} are practically zero.

The solution of the ODE converges to one of the two equilibrium curves  depending on the initial conditions.
Whereas eqs.~\eqref{eq:two-grain-nou} are invariant if both masses are multiplied by the same positive constant,
this scaling affects the initial conditions. The solution of the ODE system
with conditions $\mxo, \myo$ can thus lead to a different equilibrium regime than the solution with initial
conditions $\lambda\, \mxo, \lambda\,\myo$.

\begin{figure}
\centering
\subfigure[a]{\includegraphics[width=0.45\linewidth]{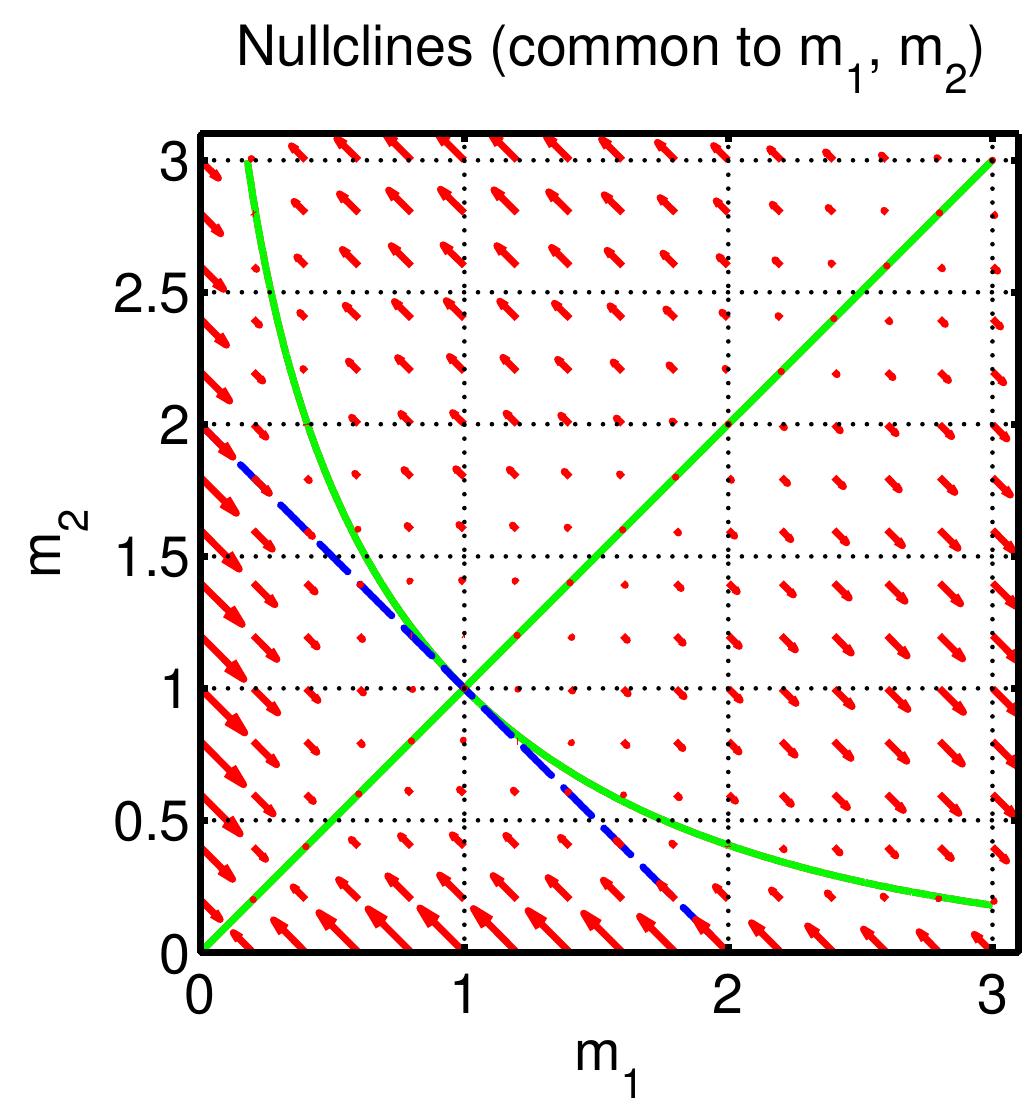}\label{fig:nullclines-a}}
\subfigure[b]{\includegraphics[width=0.45\linewidth]{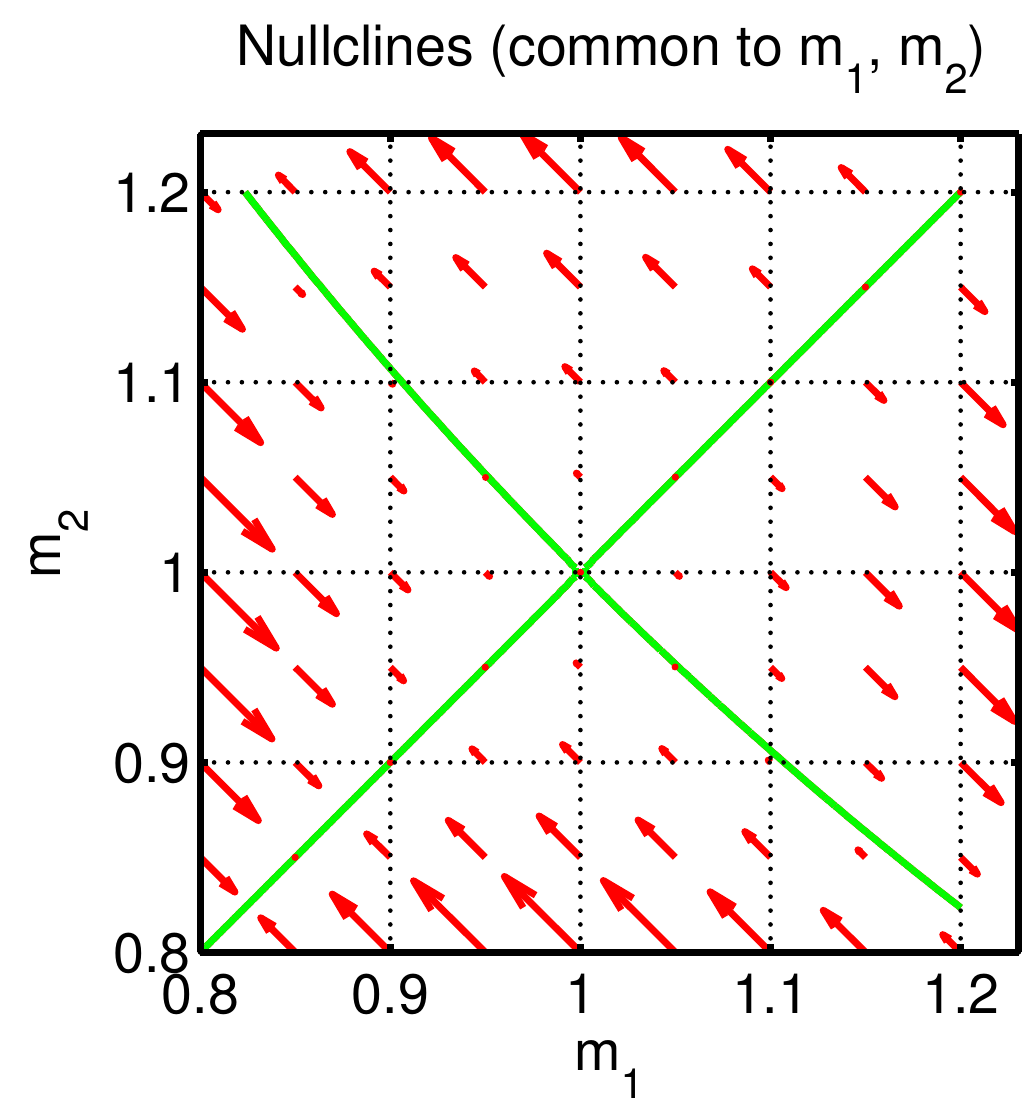}\label{fig:nullclines-b}}
 \caption{Phase plane plot showing the nullcline curves (solid lines, green online) and the mass transfer rate vectors
 for the ODE system~\eqref{eq:two-grain-nou}. (a) Diagram for mass values in the interval $[0, 3]$.
 States below the straight line $\my = 2 - \mx$  (dashes, blue online) are attracted to the equipartition nullcline,
 whereas for states above $\my = 2 - \mx$  the approach to the equipartition equilibrium is arrested by
  the separation nullcline. (b)  Detail around the point of nullclines intersection  $(1,1)$.}
 \label{fig:nullclines}
\end{figure}

The nullcline curves are displayed in Fig.~\ref{fig:nullclines}. The straight line corresponds to the
equipartition equilibrium, whereas the curvilinear trace corresponds to the growth-decay equilibrium. The two
lines intersect at point $(1, 1)$. The gradient vector $(\dot\mx(t), \dot\my(t))$ representing the
mass rate of change is also shown on these plots.
The equipartition equilibrium is stable (unstable)
for all the points that are below, i.e., to the left (above, i.e., to the right) of the
separation nullcline.
States lying below the straight line $\my = 2 - \mx$  (see Fig.~\ref{fig:nullclines-a}) are in the diffusive regime,
and they are attracted to the equipartition nullcline.
On the other hand, states below the separation nullcline and
above the line $\my = 2 - \mx$  are in the diffusive regime, but  their approach to the equipartition equilibrium is arrested at
  the mass separation nullcline.


\section{Dynamic regimes}
\label{sec:dynamic}

Below, we investigate the dynamic regimes of the two-grain system
using a theoretical analysis which is valid for certain combinations of initial conditions.
We also integrate the ODE system numerically by means of the fourth-order Runge-Kutta scheme~\cite{Press97} based on the
{\sc Matlab} ODE toolbox  function {\it ode45}, in order to determine the equilibrium state
 for the entire parameter space.

\subsection{Diffusive Regime}

These relations are preserved during the evolution of the ODE system. The latter is maintained
due to mass conservation.
The mass transfer rate remains finite and non-zero, because
 the two roots of the  transfer rate balance equation
$ f_{1,2}(t)= f_{2,1}(t)$  satisfy $m^{\ast}_{1} + m^{\ast}_{2} > 1$ (as shown in Fig.~\ref{fig:roots-rate})
for all possible transition rate levels $c$, whereas for the given initial conditions $\mx(t) + \my(t) <1$
for all $t$.

Assuming $\mx(0), \my(0) \ll 1$, we can approximate the exponential functions with one,
and the ODE system of eqs.~\eqref{eq:two-grain-nou} is approximated by the linearized equations

\begin{subequations}
\label{eq:two-grain-dif}
\begin{align}
\frac{d\lmx(t)}{dt} &  = \lmy(t) - \lmx(t),
\\
\frac{d\lmy(t)}{dt} &  = \lmx(t) - \lmy(t).
\end{align}
\end{subequations}

Eqs.~\eqref{eq:two-grain-dif} conserve mass, as we can see  by adding their respective sides.
By subtracting the left and right hand sides, respectively, of eqs.~\eqref{eq:two-grain-dif},
it follows that the \emph{mass difference,} $m_{d}(t) = \lmx(t) - \lmy(t)$, satisfies
the  ODE

\begin{equation*}
\frac{d m_{d}(t)}{dt} = - 2\,m_{d}(t),
\end{equation*}

\noi with initial condition $m_{d}(0) = \mx(0) - \my(0)$. The above equation is solved by $m_{d}(t) = m_{d}(0) \, \e^{-2 t}.$
The respective solutions  for $m_{1}(t)$ and $m_{2}(t)$ are given by the exponential functions

\begin{subequations}
\label{eq:two-grain-dif-sol}
\begin{align}
\lmx(t) & = \mx(0) \left( \frac{ 1 + \e^{-2t}}{2}\right)  +  \my(0) \left(\frac{ 1 - \e^{-2t}}{2}\right)
\\
\lmy(t) & = \mx(0) \left( \frac{ 1 - \e^{-2t}}{2} \right) +  \my(0) \left(\frac{ 1 + \e^{-2t}}{2}\right).
\end{align}
\end{subequations}

The linear approximation remains valid as $t$ increases, because mass conservation implies that
$\mx(t)$ and $\my(t)$ are bounded from above by $\mx(0) + \my(0) <1$.
The asymptotic limit of the linearized diffusive solution lies on the equipartition nullcline.

Let us recall that activation parameter $u$ is absorbed as a multiplicative
factor in the initial conditions.
Hence, lower values of $u$ tend to bring the system closer to the diffusive regime, because they
effectively reduce the ``renormalized'' initial conditions.
Since the activation parameter is inversely proportional to temperature, according to the
discussion in Section~\ref{ssec:nl-mem},
 the diffusive regime is favored by higher temperature.

\begin{figure}
\includegraphics[width=0.80\linewidth]{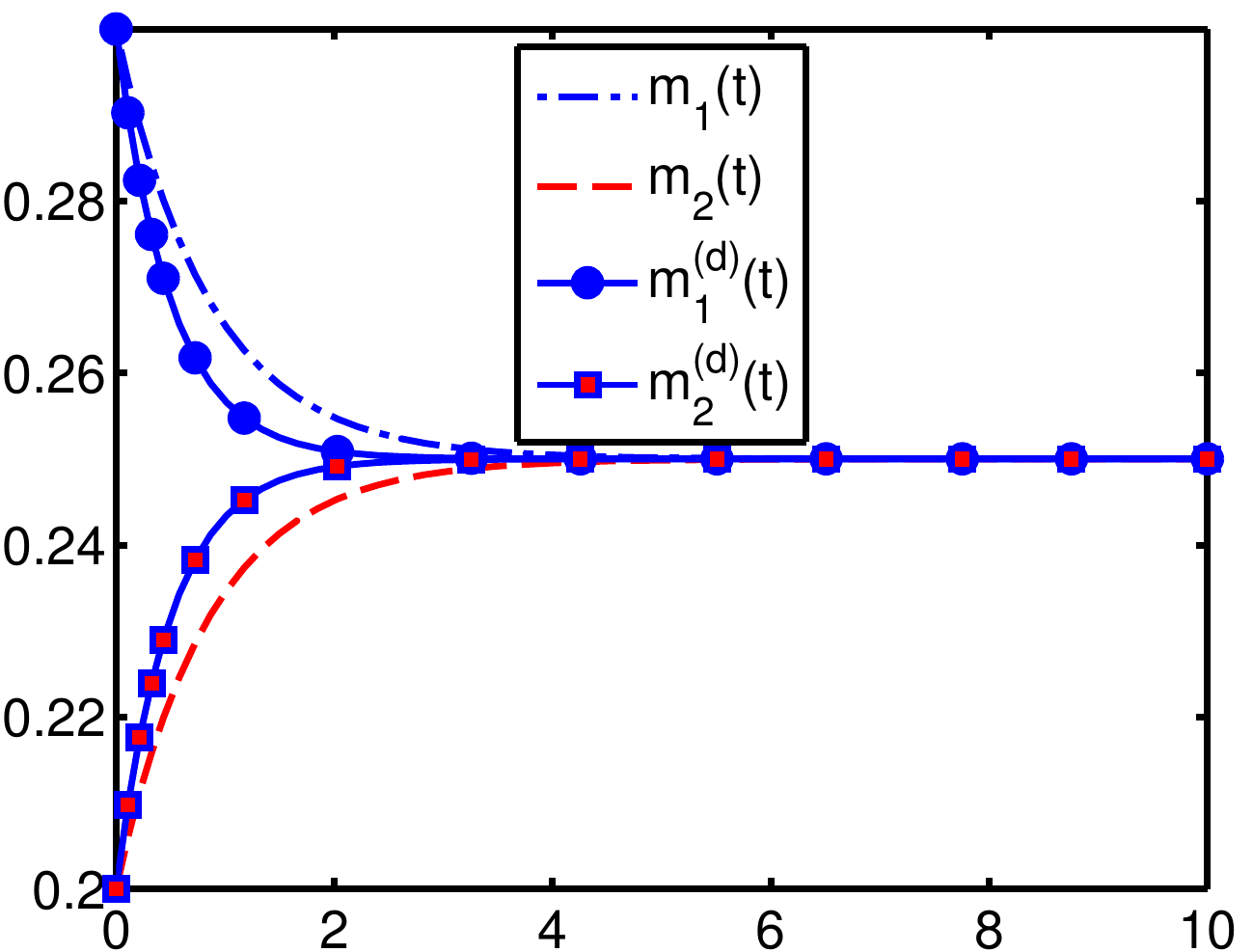}
 \caption{Numerical solutions (dashed and dash-dot lines) $\mx(t), \my(t)$  of the ODE system of eqs.~\eqref{eq:two-grain-nou} with initial conditions
 $\mxo = 0.3$ and $\myo = 0.2$ versus the explicit solutions $\lmx(t)$
 (continuous line and circles) and  $\lmy(t)$ (continuous line and squares) of the linearized equations~\eqref{eq:two-grain-dif-sol}.}\label{fig:diffusion_sol_1}
\end{figure}

In Fig.~\ref{fig:diffusion_sol_1} we compare the numerical solutions of eqs.~\eqref{eq:two-grain-nou}
with the explicit solutions of the linearized  approximation eqs.~\eqref{eq:two-grain-dif}.
The initial conditions are given by $\mxo = 0.3$ and $\myo = 0.2$, which do not strictly satisfy the
linearization conditions $\mxo\ll1, \myo\ll1$.
The numerical solution is calculated with the  explicit fourth-order Runge-Kutta
(4,5) method using a relative error tolerance of $10^{-4}$ and an absolute error tolerance of
$10^{-5}$. There is general
agreement between the two solutions, both of which converge to the same equipartition point.  The linear approximation, however,  converges
faster to equilibrium than the numerical solution. This is due to overestimation of the magnitude of mass transfer rates
in the linear approximation. On the other hand, if we use as initial conditions
 $\mxo = 0.03$ and $\myo = 0.02$ (not shown) the agreement between the numerical solution and the
 linearized approximation is excellent.

 \subsection{Growth-Decay Regime}
 \label{ssec:growth-decay}

In analogy with the preceding section, let us assume that $\mxo \neq \myo$
and that the point $(\mxo, \myo)$ is above the curvilinear separation nullcline.
Then,  the system is
in the growth-decay regime according to Fig.~\ref{fig:nullclines}. If
$\mx(0) \gg 1$, $ \my(0) \ll 1$
 we can approximate the system of eqs.~\eqref{eq:two-grain-nou} by means of
 the following asymmetric, mass-conserving, linearized approximation:

\begin{subequations}
\label{eq:two-grain-growth}
\begin{align}
\label{eq:two-grain-growth-a}
\frac{d\amx(t)}{dt} &  = \amy(t)
\\
\label{eq:two-grain-growth-b}
\frac{d\amy(t)}{dt} &  =  - \amy(t).
\end{align}
\end{subequations}

For  $\my(0) \gg 1$, $ \mx(0) \ll 1$ a similar system is
obtained from~\eqref{eq:two-grain-growth} by interchanging $\amx(t)$ and $\amy(t)$.
The solution of eqs.~\eqref{eq:two-grain-growth} is given by

\begin{subequations}
\label{eq:two-grain-growth-sol}
 \begin{align}
\amx(t) &  = \mxo + \myo (1 - \e^{-t}),
\\
\amy(t) &  =  \myo \e^{-t}.
\end{align}
\end{subequations}

\begin{figure}
\centering
\includegraphics[width=0.60\linewidth]{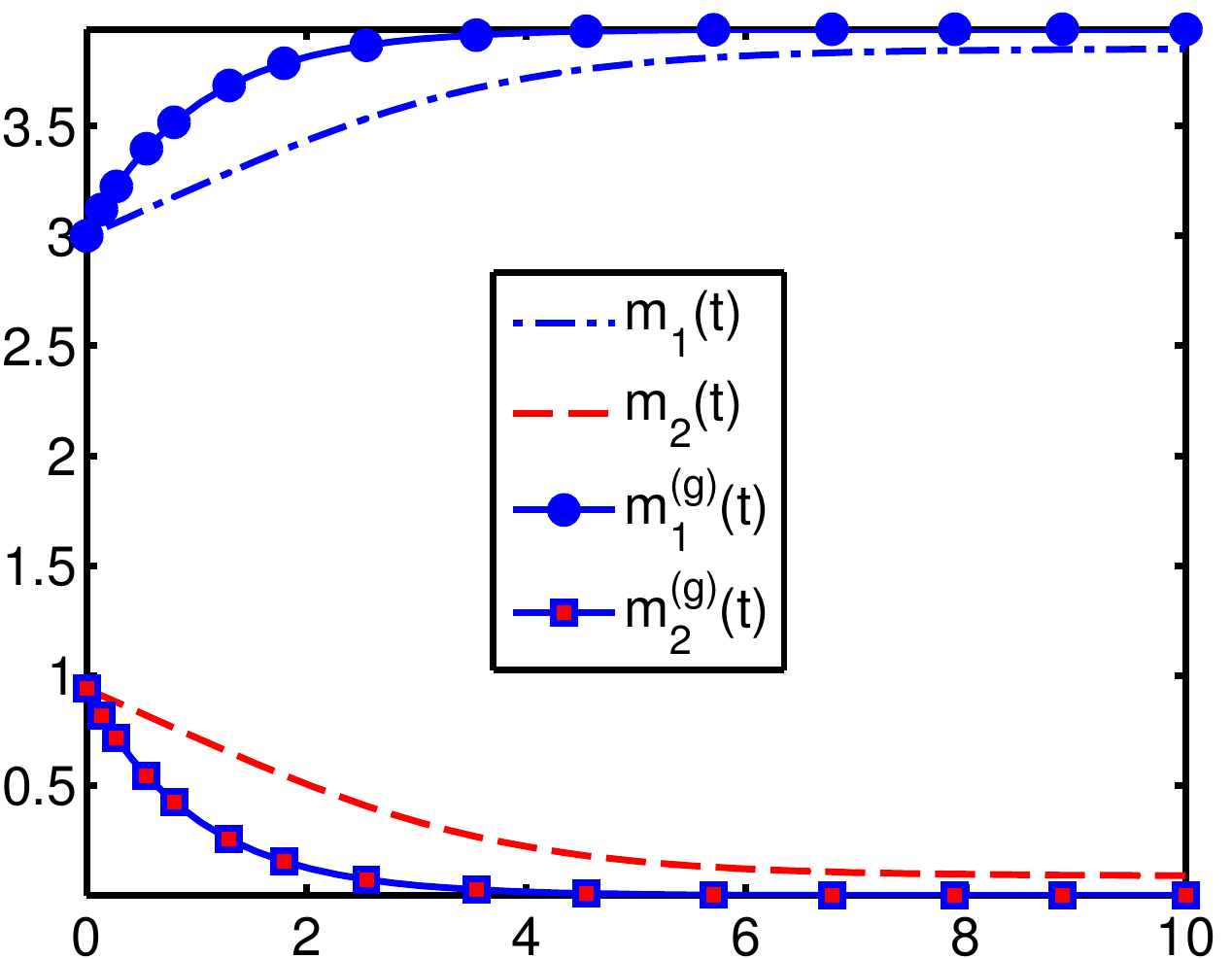}
 \caption{Numerical solutions $\mx(t), \my(t)$  of the system of eqs.~\eqref{eq:two-grain-nou} (dashed and dash-dot lines) with initial conditions
 $\mxo = 3$ and $\myo = 0.94$ compared with the explicit solution $\amx(t)$
 (continuous line and circles) and  $\amy(t)$ (continuous line and squares) given by eqs.~\eqref{eq:two-grain-growth-sol},
  of  approximate ODE system~\eqref{eq:two-grain-growth}.}\label{fig:aggregation_sol_1}
\end{figure}

The above solution predicts growth of the larger grain and decay of the smaller grain until
the former concentrates all the mass.
The solution of the ODE system~\eqref{eq:two-grain-nou}, however,
can not reach  this state, because the growth of the larger grain is trapped by
 the separation nullcline.
  In Fig.~\ref{fig:aggregation_sol_1}
we compare the numerical solution of eqs.~\eqref{eq:two-grain-nou} and the theoretical solutions of the linearized
approximation, i.e.,  eqs.~\eqref{eq:two-grain-growth}, for $\mxo = 3$ and $\myo = 0.94$.
The  difference between the two solutions
is due (i) to the  linearized approximation of the Arrhenius transition rates and
(ii) to  trapping of the nonlinear system's solution at the nullcline. The trapping effect is not
captured by linearized eqs.~\eqref{eq:two-grain-growth}.
Almost perfect agreement between the approximate and the exact solution is obtained
for  the initial condition $\mxo = 16$ and $\myo = 0.5$ (not shown herein), which is closer to the validity regime
for approximation~\eqref{eq:two-grain-growth}.

\subsection{Trapping}

\begin{figure}
\centering
\includegraphics[width=0.60\linewidth]{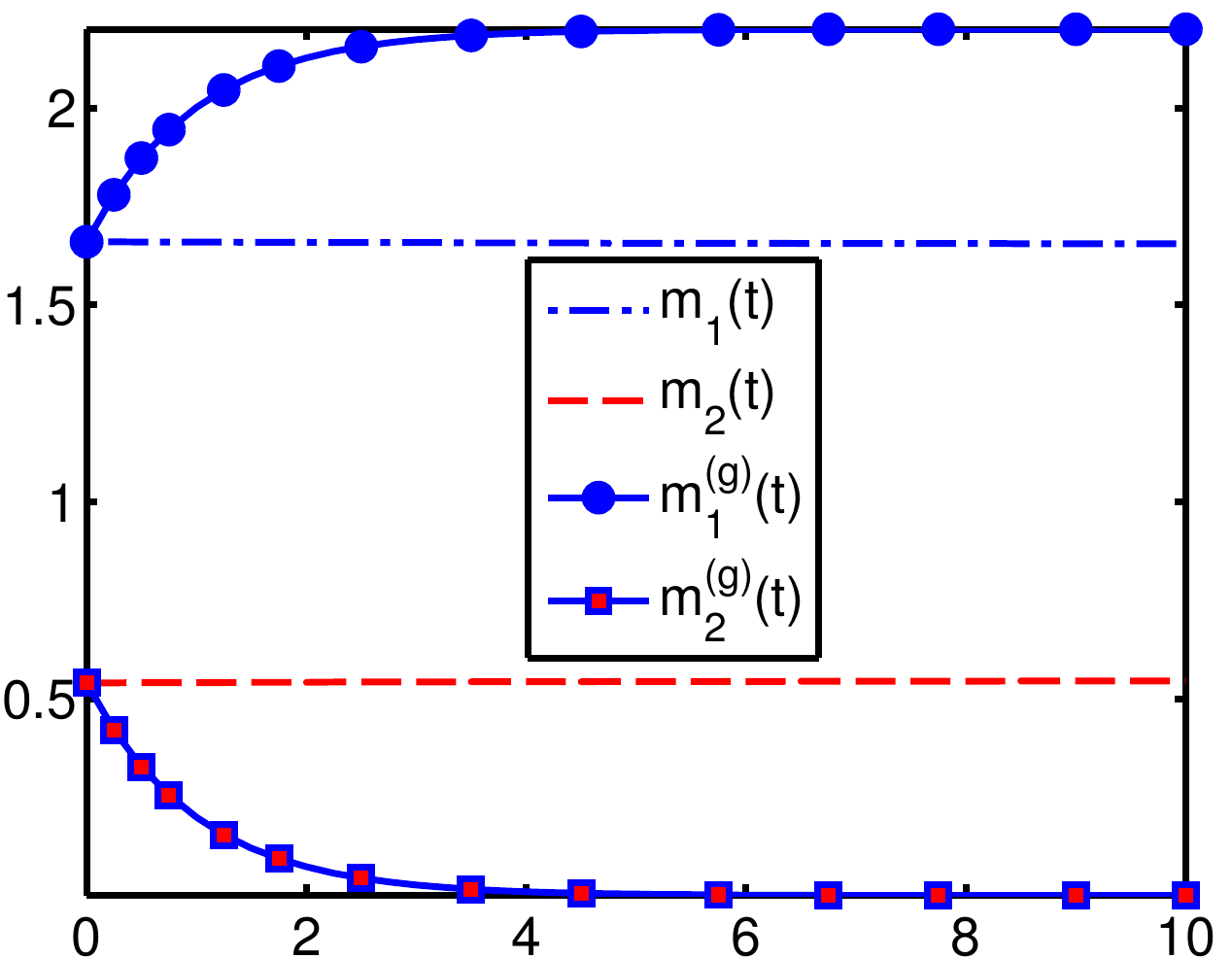}
 \caption{Numerical solutions $\mx(t), \my(t)$  of the system of eqs.~\eqref{eq:two-grain-nou}
 (dashed and dash-dot lines)  with initial conditions
 $\mxo = 1.66$ and $\myo = 0.54$ (trapped state) versus the explicit solution $\amx(t)$
 (continuous line and circles) and  $\amy(t)$ (continuous line and squares)
 given by eqs.~\eqref{eq:two-grain-growth-sol}
  of the approximate ODE system given by eqs.~\eqref{eq:two-grain-growth}.}\label{fig:trapping_sol_1}
\end{figure}

If $\mx(0), \my(0) \gg 1$ the system is essentially frozen in the initial state, because the transition rates $ f_{1,2}(t)$ and
 $f_{2,1}(t)$ are nearly zero. This also occurs for equal initial masses, i.e.,
  if $\mx(0) = \my(0) =m_0$ for all  $m_0$.
  In addition,  trapping occurs if the evolution in the diffusive regime stops at the separation nullcline.
  Trapping in the initial state is
 illustrated in Fig.~\ref{fig:trapping_sol_1}:
the initial masses,  $\mxo = 1.66$ and $\myo = 0.54$, correspond
to mass transfer rates with magnitude $\approx 10^{-3}$, because the initial point
lies approximately  on the mass separation nullcline. Hence, the system is essentially frozen in this state.
The linearized  grain coarsening approximation, eq.~\eqref{eq:two-grain-growth-sol}, however, is insensitive to this effect and predicts mass transfer
from the smaller to the larger grain. This is not surprising, since the conditions of validity for the
linearized approximation are not satisfied.

\subsection{Equilibrium Phase Diagram }
\label{ssec:phase-diagram}
\begin{figure}[ht]
\begin{center}
\subfigure[a]{\includegraphics[width=0.53\linewidth]{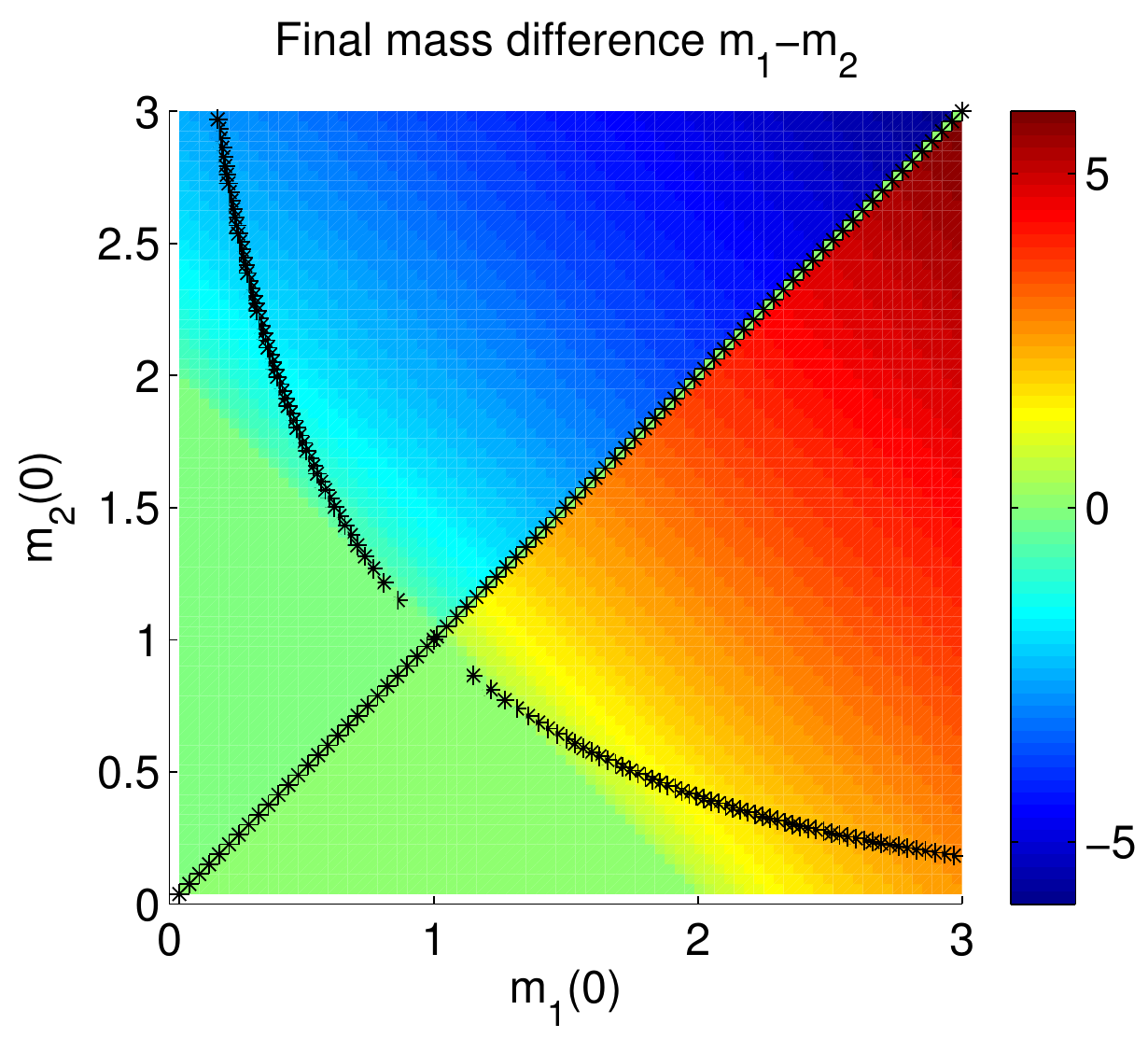}
 \label{fig:m1-m2-a}}
\subfigure[b]{\includegraphics[width=0.43\linewidth]{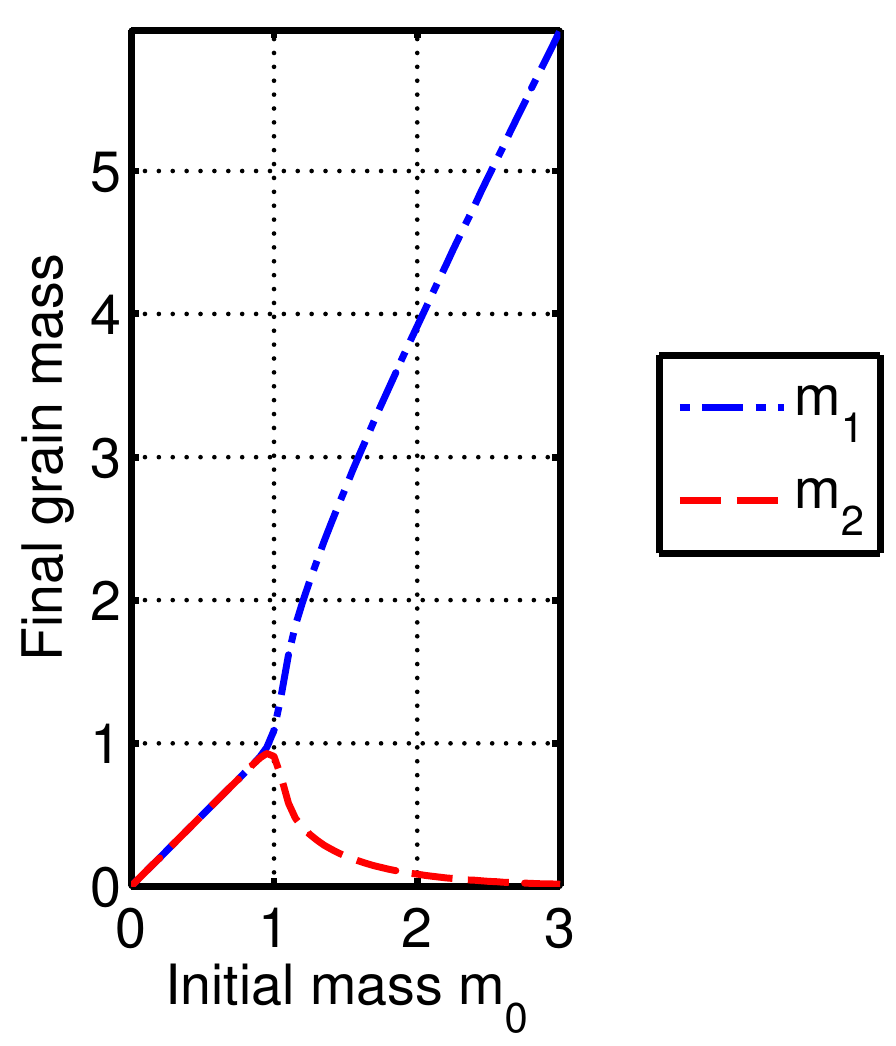}
\label{fig:m1-m2-f}}
\end{center}
 \caption{(a) Grain mass difference $m_{d}(T) = \mx(T) - \my(T)$  at ``equilibrium''
 $(T=40)$  as a function of the initial conditions in $[0, 3] \times [0, 3]$. The
 horizontal and vertical axes denote the initial grain masses, $m_{1}(0)$ and $m_{2}(0)$, respectively.
 The shading (color online) denotes the mass difference $m_{d}(T)$
 at equilibrium at the nodes of an $80 \times 80$ grid.
 Equilibrium values are based on numerical solutions of ODE system~\eqref{eq:two-grain-nou} until $|d m_{d}/d t| < 2\times 10^{-4}$.
 (b) Plot of the final grain masses at $T=40$ versus  $m_{0}$ where the latter defined the initial conditions
 $\mxo= m_{0}+ \delta m, \myo=m_{0} - \delta m$, using $\delta m=0.1$. }
 \label{fig:m1-m2}
\end{figure}

The phase diagram of Fig.~\ref{fig:m1-m2-a} illustrates  the equilibrium state of
eqs.~\eqref{eq:two-grain-nou}  over a subset  of the parameter plane $(\mxo,\myo)$.
Since the total mass is conserved, the crucial state variable is the
mass difference, which obeys the following equation:

\begin{equation}
\label{eq:mass-diff}
\frac{d m_{d}(t)}{d t} = 2 \left[  f_{2,1}(t) - f_{1,2}(t) \right].
\end{equation}

The initial mass difference is
(i) amplified in the growth-decay regime, (ii) reduced in the diffusive regime, or
(iii) maintained in the trapped state.
The  equilibrium mass difference, $m_{d,\rm{eq}} =\lim_{t \ra \infty} m_{d}(t)$,
is derived from the numerical solution of eqs.~\eqref{eq:two-grain-nou}.  The equilibrium is determined
by requiring that $|d m_{d}/d t| < 2\times 10^{-4}$ for all $\mx(0), \my(0)$ examined.
As Fig.~\ref{fig:m1-m2}(a) shows, in the diffusive regime (lower left),
 the mass difference tends to zero. In the
growth-decay regime (upper right), the larger  grain concentrates most of the mass.
The  curvilinear trace of the separation nullcline and the straight-line equipartition nullcline are marked by
``star'' pointers.

The discontinuous
 change of the equilibrium mass difference near the main diagonal
 (upper right corner) is  triggered by a small change in the initial conditions: the system moves
  from the equipartition point (static regime) which extends along the main diagonal to
large mass differences (negative above and positive below the diagonal) in the adjacent growth-decay regime.
his is illustrated in Fig.~\ref{fig:m1-m2}(b), which displays the final masses of the two grains assuming that the initial conditions are
incrementally different, i.e.,
$\mxo = m_{0} + \delta m$ and  $\myo = m_{0} - \delta m$, where $\delta m=0.1$.

\subsection{Grain Mass Evolution }
\label{ssec:num-evolution}
\begin{figure}
\centering
\includegraphics[width=0.8\linewidth]{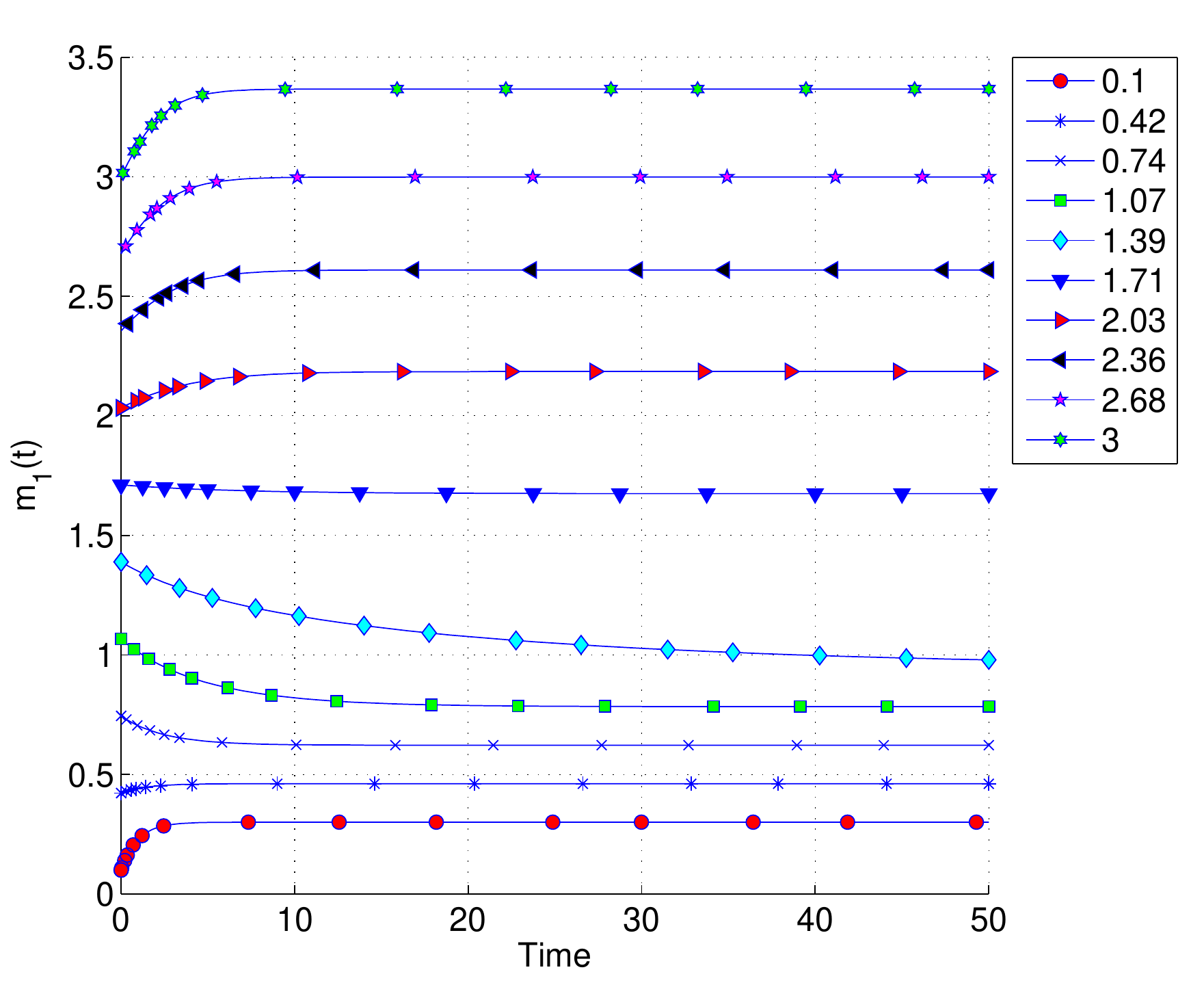}
 \caption{Evolution of $\mx(t)$ based on numerical solutions of eqs.~\eqref{eq:two-grain-nou} for ten
 different values of $\mx(0)$ (shown in the legend truncated to second decimal) and $\my(0)=0.5$.  }
 \label{fig:m1-t}
\end{figure}

\begin{figure}
\centering
\includegraphics[width=0.8\linewidth]{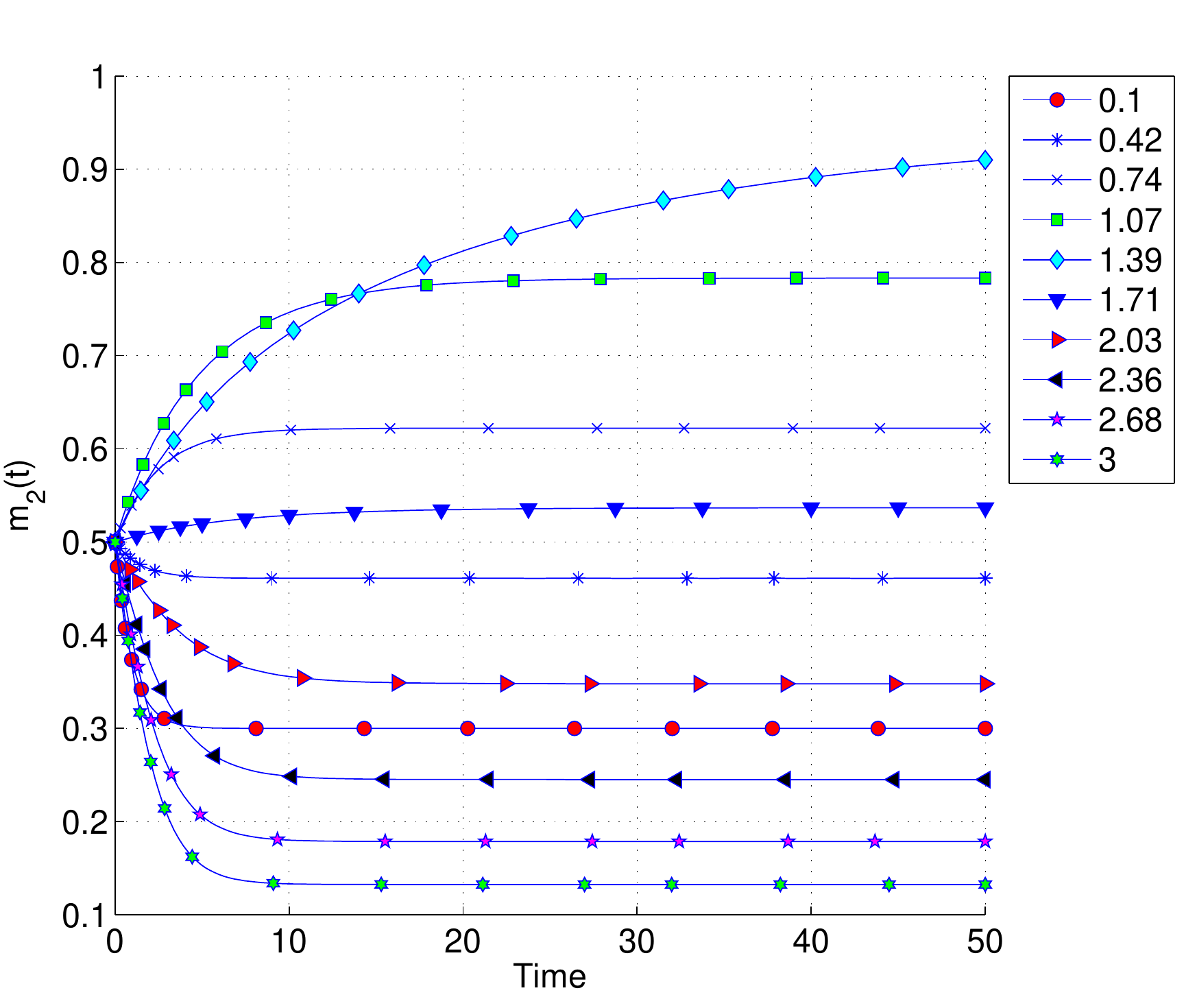}
 \caption{Evolution of $\my(t)$ based on numerical solutions of eqs.~\eqref{eq:two-grain-nou}  for ten
 different values of $\mx(0)$ (shown in the legend truncated to second decimal) and $\my(0)=0.5$. }
 \label{fig:m2-t}
\end{figure}

We plot the evolution of the masses $\mx(t)$ and $\my(t)$ over time in Figs.~\ref{fig:m1-t} and~\ref{fig:m2-t}, respectively.
The plots are generated for fixed $\my(0)=0.5$ whereas $\mx(0)$ varies between 0 and 3. As shown in Fig.~\ref{fig:m1-t},
for $\mx(0) < 0.5$, $\mx(t)$ increases with $t$ marking the diffusion of mass from $\my(t)$ to $\mx(t)$. The process is
much slower for $\mx(0) \approx 0.42$, because the system is close to the equilibrium point $(0.5, 0.5)$. For $ 0.5 <\mx(0) \lessapprox 1.71$,
$\mx(t)$ declines with $t$ due to mass diffusion  towards the initially smaller grain.
In contrast, for $\mx(0) \ge 2.03$, $\mx(t)$ grows with $t$ again, marking the entrance in the growth-decay regime.
The dependence of $\my(t)$ is diffusive for $\mx(0) \leq 1.71$; $\my(t)$ declines for $\mx(0) < \my(0)$ and
grows for $\mx(0) > \my(0)$. For $\mx(0) \ge 2.03$ the system is in the growth-decay regime, and since
$\mx(0) > \my(0)$, $\my(t)$ decays by transferring mass  to $\mx(t)$.

\begin{figure}
\centering
\includegraphics[width=0.9\linewidth]{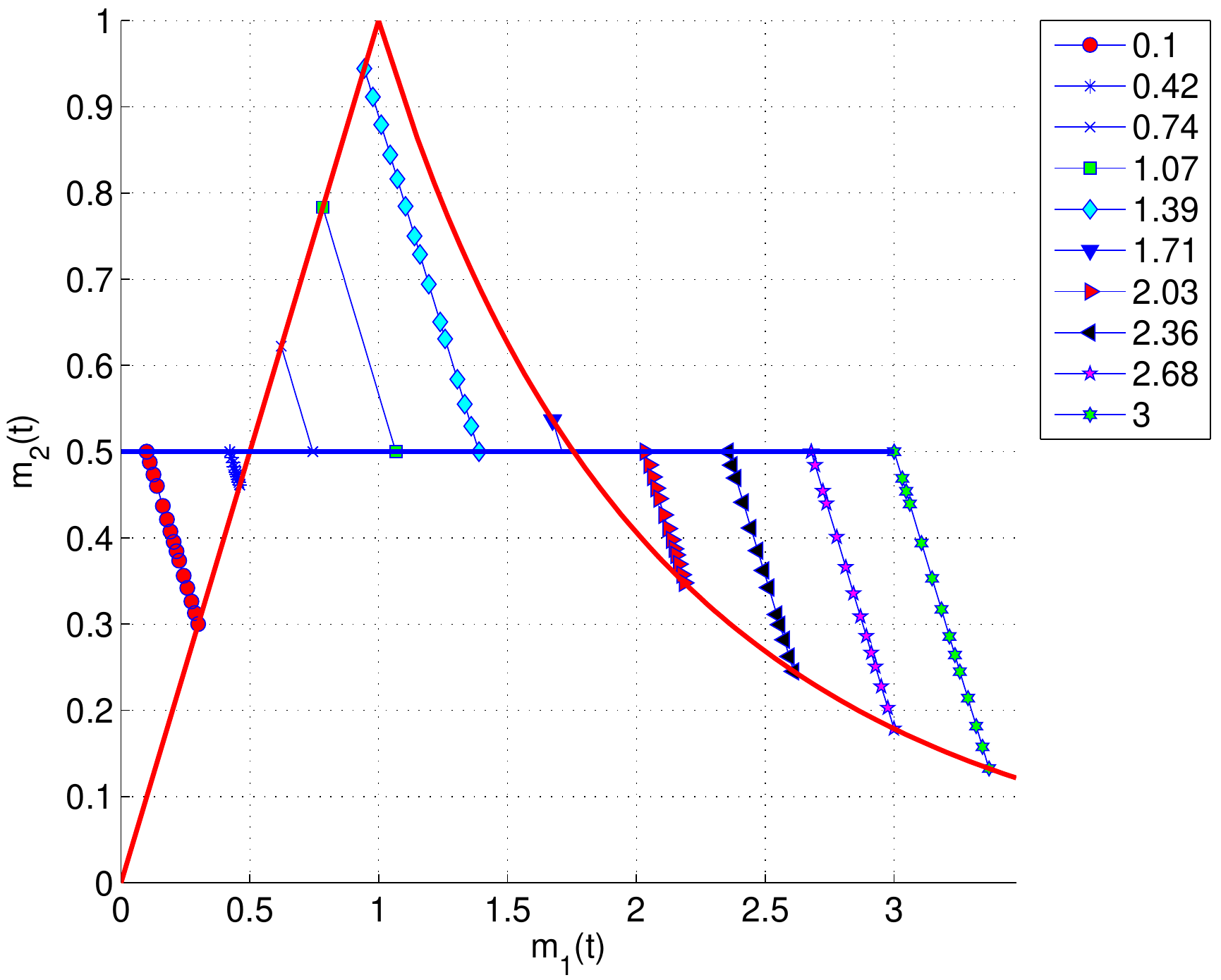}
 \caption{Trajectories  of  grain masses $\mx(t), \my(t)$ with $\my(0)=0.5$ and for ten
 different values of $\mx(0)$ (shown in the legend truncated to second decimal) from
 $t=0$ to $t=200$. The solid
 straight and curved lines (red online) represent the equipartition and
 separation nullclines respectively. The horizontal line intercepting the vertical axis at $0.5$ (blue online) marks the initial condition $\my(0)$. }
 \label{fig:m1t-m2t}
\end{figure}

The trapping effect is not transparent in Figs.~\ref{fig:m1-t}-\ref{fig:m2-t}. To better illustrate it,
 we plot the trajectories $\left(  \mx(t), \my(t) \right)$ in Fig.~\ref{fig:m1t-m2t}.
In the same plot we include the equipartition nullcline along
  the diagonal and the curvilinear separation nullcline.
  Due to the mass conservation constraint, the trajectories obey the linear equation  $\my(t) = \mx(0) + \my(0) - \mx(t)$.
  Hence,  in spite of the fact that $\mx(t)$ and $\my(t$) satisfy first-order ODEs that are
  nonlinear and cannot be solved analytically, the evolving masses satisfy a simple linear
  relation. For the  two lowest values of $\mx(0)$, the
system behaves diffusively, with $\mx(t)$ growing and $\my(t)$ shrinking.
The next three values of $\mx(0)$  also lead to diffusive behavior, with $\mx(t)$ declining and $\my(t)$ growing.
All of the first five trajectories terminate on the equipartition nullclline.
The next value,
$\mx(0) = 1.71$, also leads to diffusive behavior which is arrested at the separation nullcline.
The trapping of diffusive solutions by the separation nullcline is not captured by the solution
of the linearized approximation
in the diffusive regime given by~\eqref{eq:two-grain-dif-sol}.
Finally, values of $\mx(0) \ge 2.03$ lead to growth of the larger grain,
 which is arrested at the growth-decay equilibrium curve.

\section{Stochastic Transition Rates}
\label{sec:noise}
Trapping can be undesirable in certain cases, because it impedes either the diffusion or the grain coarsening process.
One way  to escape trapping is by adding noise to the ODE system~\eqref{eq:two-grain-nou},
leading to

\begin{subequations}
\label{eq:two-grain-nou-noise}
\begin{align}
\frac{d\mx(t)}{dt} &  = f_{2,1}(t) - f_{1,2}(t) - \eta(t)
\\
\frac{d\my(t)}{dt} &  =  f_{1,2}(t) - f_{2,1}(t) + \eta(t),
\end{align}
\end{subequations}

\noindent where $\eta(t)$ is Gaussian white noise with zero mean random process, i.e.,  $\langle \eta(t) \rangle = 0$,  and
correlation function
$\langle \eta(t)\, \eta(t') \rangle = \vars \delta(t-t')$, $\vars$ being the noise variance.
Eqs.~\eqref{eq:two-grain-nou-noise}
conserve the total mass at all times due to the opposite signs of the noise terms.

If at  $t>0$, the  transfer rate balance is reached, it will almost surely be perturbed by stochastic fluctuations.
The mass difference satisfies the equation

\begin{equation}
\label{eq:mass-diff-noise}
\frac{d m_{d}(t)}{d t} = 2 \left[  f_{2,1}(t) - f_{1,2}(t) \right] + 2 \, \eta(t).
\end{equation}

\subsection{Special cases}

If $f_{2,1}(t), f_{1,2}(t) \approx 0$, $m_{d}(t)$ is essentially driven by the white noise process.
Hence, $m_{d}(t)$ becomes a Wiener process  which
describes the classical Brownian motion~\cite{Zwanzig01}.

In the region of the growth-decay regime where the linearized approximation holds
(see Section~\ref{ssec:growth-decay}),
the mass of the second grain based on eqs.~\eqref{eq:two-grain-growth} satisfies the equation

\begin{equation}
\label{eq:mass-diff-growth-noise}
\frac{d \amy(t)}{d t} = - \amy(t) + \, \eta(t),
\end{equation}

\noi which represents the \emph{Ornstein-Uhlenbeck}  process~\cite{UO30}.
The solution of the latter is given by the following stochastic integral:
\[
\amy(t) = \amy(0) \, \e^{-t}  + \int_{0}^{t} \e^{-(t-u)} \, dW(u),
\]
where $W(t)$ is the Wiener process~\cite{Gillespie96}.
Using the conservation of mass in the system, the mass of the first grain is given by
$\mx(0)+\my(0) - \amy(t)$, which is also an  Ornstein-Uhlenbeck process.

\subsection{Numerical simulations}
Given the lack of explicit solutions,
one can numerically integrate eqs.~\eqref{eq:two-grain-nou-noise} using the Euler-Maruyama scheme which employs
the  updating rule
\begin{subequations}
\label{eq:two-grain-nou-noise-integ}
\begin{align}
\mx(t + \de t) &  = \mx(t) + \left[ f_{2,1}(t) - f_{1,2}(t)\right] \de t - \sigma \,e(t) \sqrt{\de t},
\\
\my(t + \de t) &  = \my(t) + \left[ f_{1,2}(t) - f_{2,1}(t)\right] \de t + \sigma \,e(t) \sqrt{\de t}
,
\end{align}
\end{subequations}
where $\sigma$ is the noise standard deviation and $e(t)$ is a realization
of the Gaussian white noise process $N(0,1)$. The above scheme
is known to produce accurate results only if $\de t \ll 1$.

The equilibrium (long-time limit)
of the noisy system is not  a stationary state, since the mass
transition rates fluctuate around zero. The  noise has the most impact if the initial mass values
are in close proximity to the nullclines or between the growth decay nullcline and
 the line $\mxo + \myo =2$. As discussed in Section~\ref{sec:equil}, in the latter region
 the system is in the diffusive regime, but the mass trajectories are arrested by the separation nullcline.

\begin{figure}[ht]
\begin{center}
\subfigure[$\mxo = 2.5$, $\myo=0.49$]{\includegraphics[width=0.45\linewidth]{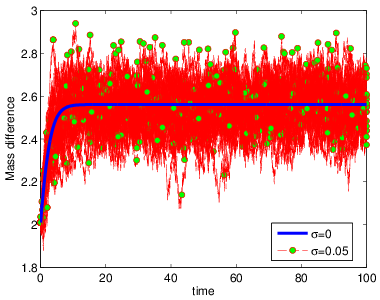}\label{fig:noise-a}}
\subfigure[$\mxo = 2$, $\myo=0.2$]{\includegraphics[width=0.45\linewidth]{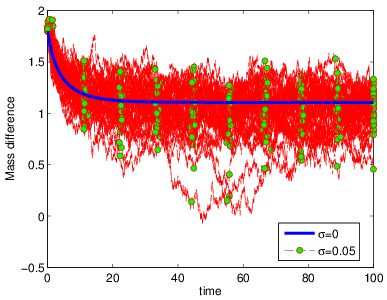}\label{fig:noise-b}}
\subfigure[$\mxo = 2.5$, $\myo=2.49$]{\includegraphics[width=0.45\linewidth]{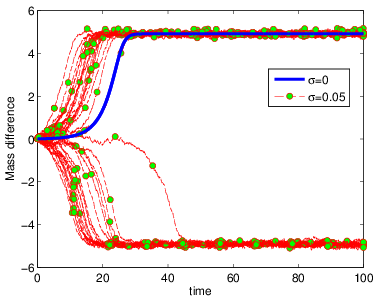}\label{fig:noise-c}}
\subfigure[$\mxo = 6.5$, $\myo=6.49$]{\includegraphics[width=0.45\linewidth]{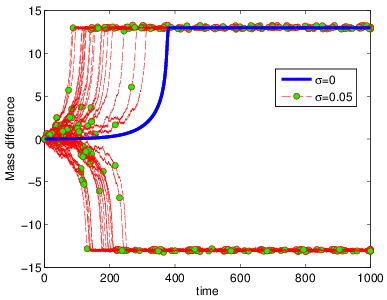}\label{fig:noise-d}}
\end{center}
 \caption{Plot of the grain mass difference $m_{d}(t)$ versus time based on the solution of
 ODE system~\eqref{eq:two-grain-nou} for zero noise and system~\eqref{eq:two-grain-nou-noise} with
 $\sigma=0.05$.  The noise-free system is solved using the Euler scheme, and the noisy system using
 the Euler-Maruyama scheme given by~\eqref{eq:two-grain-nou-noise-integ}.
 We use a time step $\delta t = 10^{-4}$ and a total number of $N=10^6$ steps
 except  $\mxo = 6.5$, $\myo=6.49$ for which $N=10^7$ steps are used. The
 mass differences involve samples taken every $100$ time steps.  Forty simulations are used in
 the noisy case; they are marked by circles (green online) and dashed lines (red online).
 The thick continuous line (blue online) represents the noise-free solution. }
  \label{fig:ode-noise}
\end{figure}

 We calculate the grain mass difference $m_{d}(t)$ versus time as obtained from
 the solution of
 system~\eqref{eq:two-grain-nou} for zero noise and of system~\eqref{eq:two-grain-nou-noise}
 in the noisy case with
 $\sigma=0.05$.  The noise-free system is solved using the Euler scheme, whereas the noisy system uses
 the Euler-Maruyama scheme given by~\eqref{eq:two-grain-nou-noise-integ}.
 We use a time step $\delta t = 10^{-4}$ and a total of $N=10^6$ steps. In the case
   $\mxo = 6.5$ and $\myo=6.49$, we use $N=10^7$ steps to capture the slow approach to equilibrium. The
 mass differences shown in Fig.~\ref{fig:ode-noise}
 represent samples taken every $100$ time steps.

The plots in Fig.~\ref{fig:noise-a}
 correspond to evolution of the system in the growth-decay regime, whereas
 those in Fig.~\ref{fig:noise-b} correspond to the diffusive regime. In both cases the
 noise leads to significant dispersion, but it does not reverse the deterministic trend.
 Figs.~\ref{fig:noise-c}-\ref{fig:noise-d} correspond to initial conditions that
 are very close to the equipartition nullcline, but in the region of the phase diagram
 where the equipartition nullcline cuts through the growth-decay regime.
 These states evolve towards the growth-decay equilibrium in the
 noise-free system. In the noisy system, on the other hand, the approach to the fluctuating equilibrium is
 in general faster, because the stochastic forces drive the system away from the equipartition nullcline.
 In  some of the simulations shown in Figs.~\ref{fig:noise-c}-\ref{fig:noise-d},
  the initially smaller grain   grows at the expense of the
 bigger grain. For the initial state  $\mxo = 6.5$, $\myo=6.49$ which is close to the equipartition nullcline,
 the approach to the equilibrium is very slow due
 to the small value of $|f_{1,2}(t) - f_{2,1}(t)|$. Thus, significantly longer runs are necessary to
 approach the equilibrium.

\section{$N$-Grain Mass Exchange Model}
\label{sec:N-grains}
Let us consider  a system comprising $N$ grains with masses $m_{i}(t)$, $i=1, \ldots, N$ and periodic boundary conditions, so that
$m_{N+1}(t) \equiv m_{1}(t)$ and $m_{0}(t) \equiv m_{N}(t)$.  Each grain is assumed to interact only with its left- and right-hand
nearest neighbors.
As discussed in Section~\ref{sec:two-grain}, the activation parameter can be absorbed in the initial conditions.
Nevertheless, we opt to preserve $u$ for reasons explained below.
Then,  system~\eqref{eq:mass-trans2} for the evolution of the grain masses is expressed as follows for $i=1, \ldots, N$

\begin{equation}
\label{eq:mass-transN}
\frac{d {m}_{i}(t)}{d t}   =
\e^{  -  u {m}_{i-1}(t)   }  {m}_{i-1}(t) + \e^{  -  u {m}_{i+1}(t)   }  {m}_{i+1}(t)
   -   2 \e^{  - u {m}_{i}(t) } {m}_{i}(t),
\end{equation}
subject to the initial conditions $m_{i}(t_0)$, $i=1, \ldots, N$, at time  $t_{0}$.

This system also conserves mass, as shown by adding the rates of mass change for all the grains, because
each term $\e^{  - u  \, {m}_{i}(t) } {m}_{i}(t)$ appears once with a coefficient equal to $-2$ ---in the mass evolution equation for $m_{i}(t)$---
and twice with a coefficient equal to one ---in the mass evolution equations for $m_{i-1}(t)$ and $m_{i+1}(t)$.
The conserved system mass is thus given by $m_{\rm tot}=\sum_{i=1}^{N} m_{i}(t_{0})$.

Numerical investigations show that system~\eqref{eq:mass-transN}  tends to
equipartition or the growth-decay equilibrium. The factors that determine the equilibrium involve the initial mass
distribution and the  activation parameter.
If we assume that $ u \, m_{i}(t_{0}) \ll 1$ for all $i=1, \ldots, N$, eq.~\eqref{eq:mass-transN}
is approximated by the following

 \begin{equation}
\label{eq:mass-transN-approx}
\frac{d {m}_{i}(t)}{d t}   = {m}_{i-1}(t) +  {m}_{i+1}(t)  -   2  {m}_{i}(t).
\end{equation}

If the grains are at distance $a$ from each other, the right hand side of the above equation is
a discrete approximation of the continuum limit
$\approx a^{2} \, \partial^{2} m(x,t) / \partial x^{2}$.
Hence, eq.~\eqref{eq:mass-transN-approx}  tends to the diffusion equation
$\partial {m}(x,t)/ \partial t = a^{2} \partial^{2} m(x,t) / \partial x^{2}$.
 If $u \sim \mathrm{O}(1)$ and $m_{i}(t_0)\sim \mathrm{O}(1/N)$  for
all  $i=1, \ldots, N$,  the evolution is diffusive regardless of the specific
initial distribution  $m_{i}(t_0)$.

The relaxation of the $N$-grain system is in general slower than that of the
two-grain system due to the many degrees of freedom involved.
The reason is that equilibrium is reached only if
the mass transfer rates simultaneously vanish for all the grains.
This condition is established if all the masses converge to the same value,
if the growth-decay process leads to a grain mass distribution
 that practically traps the system on the separation nullclines,
 or if the diffusion is trapped at the separation nullcline.

\begin{figure}[ht]
\begin{center}
\subfigure[$u=0.05$]{\includegraphics[width=0.45\linewidth]{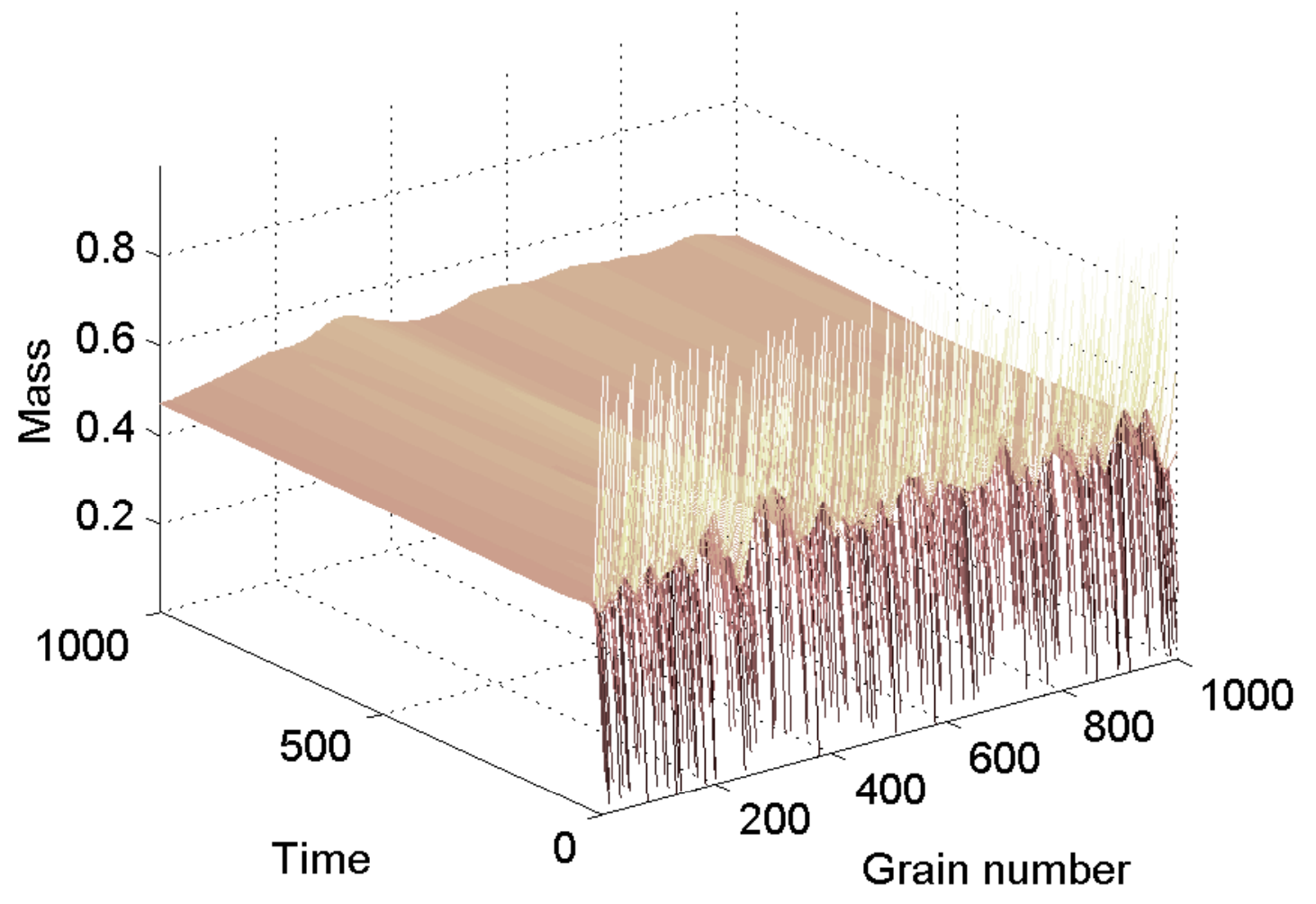}\label{fig:u-evol-a}}
\subfigure[$u=1$]{\includegraphics[width=0.45\linewidth]{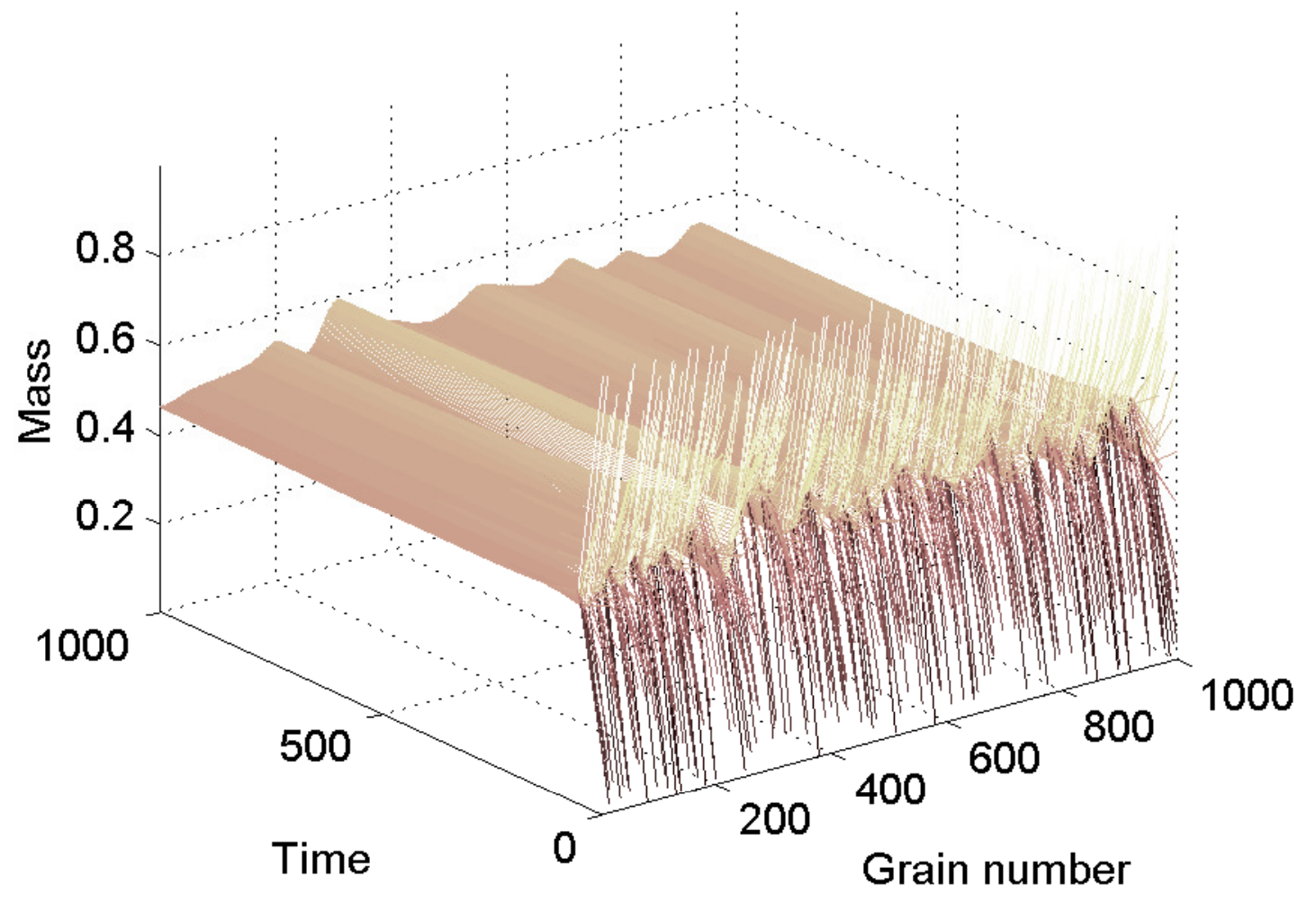}\label{fig:u-evol-b}}
\subfigure[$u = 1.5$]{\includegraphics[width=0.45\linewidth]{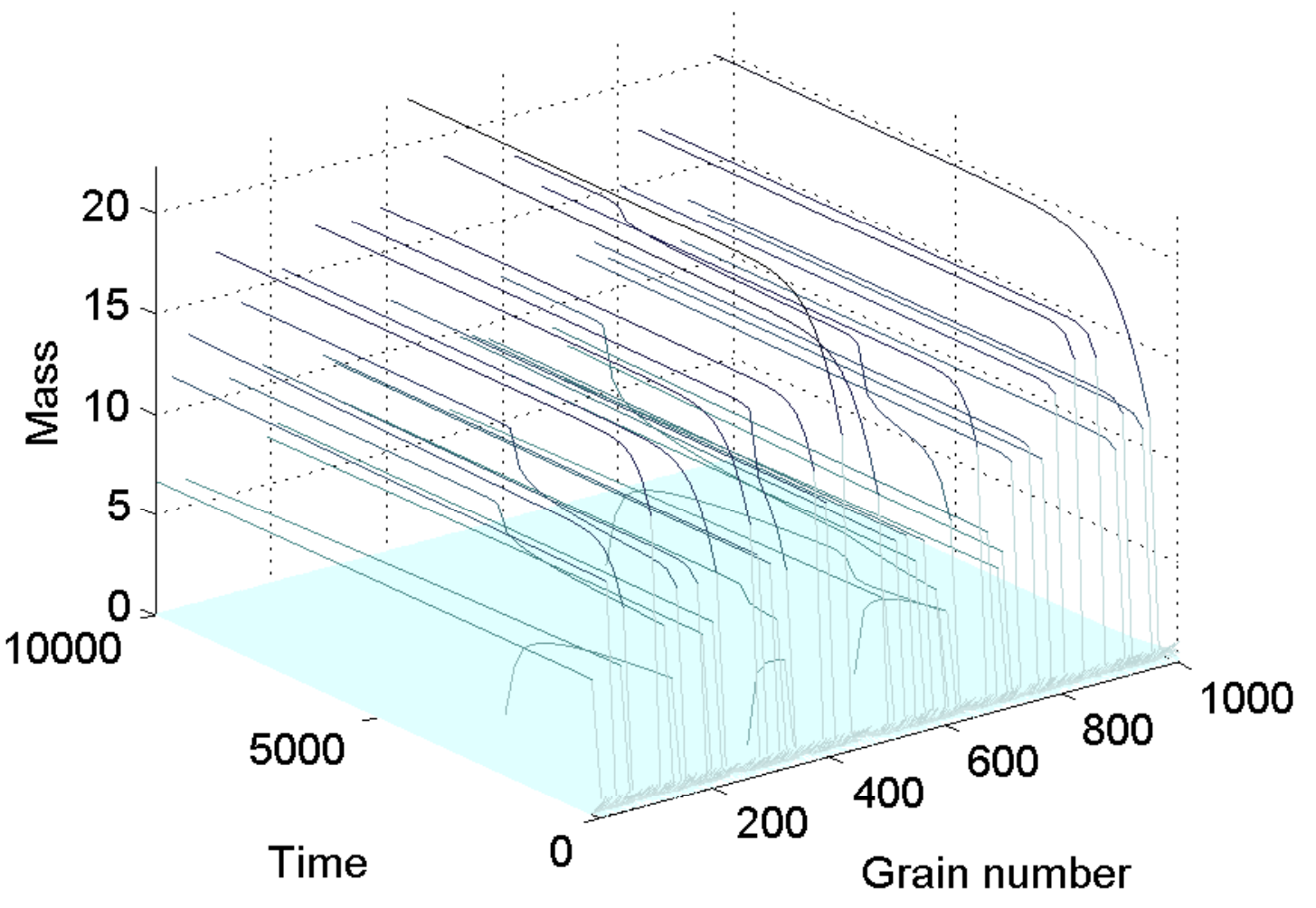}\label{fig:u-evol-c}}
\subfigure[$u =2$]{\includegraphics[width=0.45\linewidth]{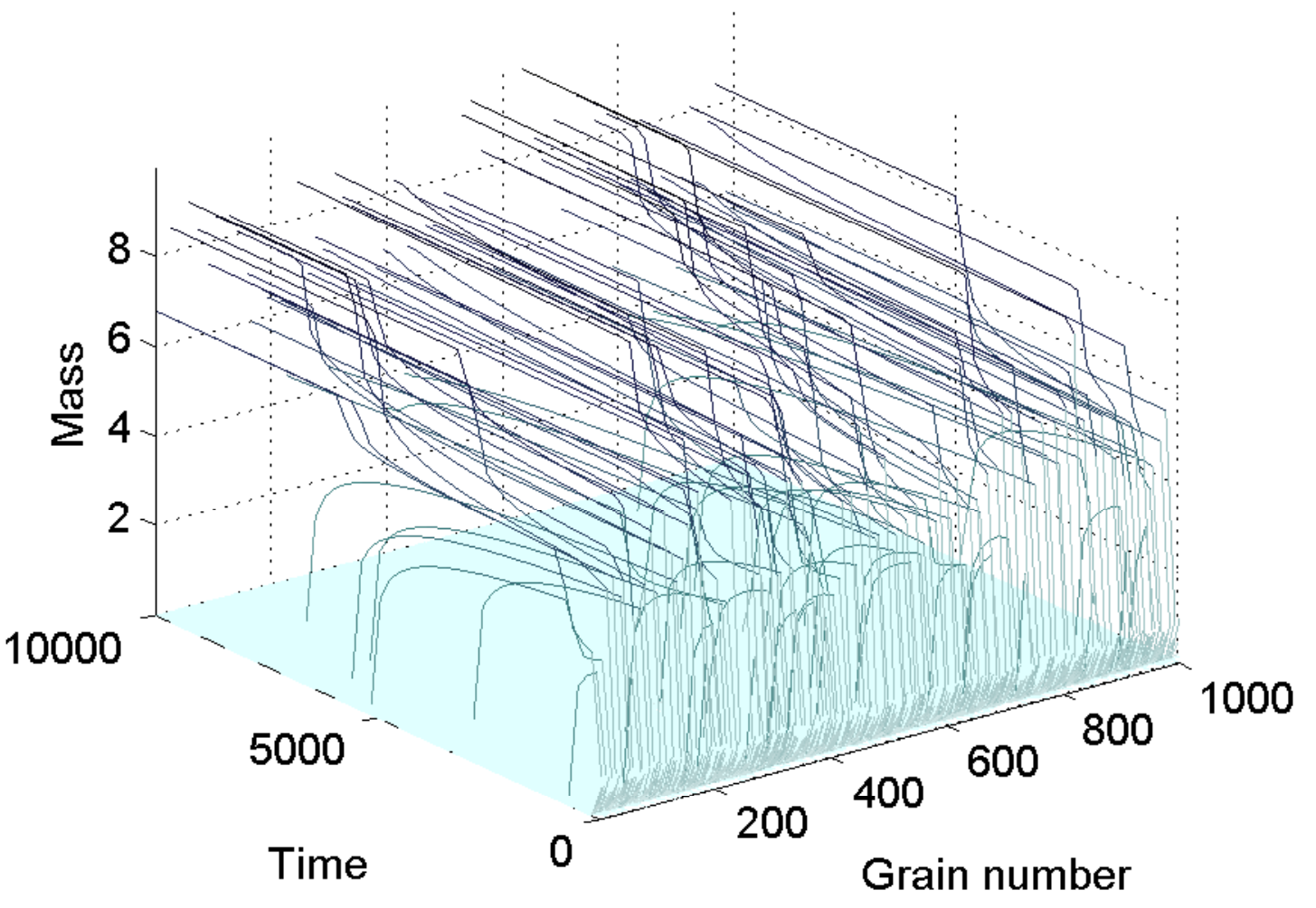}\label{fig:u-evol-d}}
\end{center}
 \caption{Evolution of a system comprising $N=10^3$ grain masses versus time for different values of $u$ based on the solution of
 ODE system~\eqref{eq:mass-transN}.  The final time is $T=10^3$ for $u \le 1$ and $T=10^4$ for $u>1$.
 The masses are sampled with a step equal to $100$. The initial mass configuration is drawn from the uniform distribution $U(0,1)$.  The vertical axis
 in all plots corresponds to grain mass. }
  \label{fig:u-evol}
\end{figure}

 We illustrate the evolution of a system containing $10^3$ grains in Fig.~\ref{fig:u-evol} for
 four different values of $u=[0.05, 1, 1.5,  2]$, assuming that the initial grain masses are drawn
 from a uniform distribution over the interval $[0, 1]$. This is equivalent to $u=1$ and an initial
 grain mass distribution in $[0, u]$.  For $u=0.05$ and $u=1$ the system tends to
 the equipartition equilibrium. At $T=1000$ mass fluctuations survive, but the grains have approached $0.5$.
 On the other hand, for $u=1.5$ and $u=2$ the system is in the growth-decay regime. The relaxation is considerably slower,
 and thus we extend the final time to $T=10^4$. Many grains quickly tend to zero mass (hence the dark shading of the
 $z=0$ plane), whereas a smaller number of grains (fewer than 100) increase their mass; the mass evolution of these grains is shown
 by the thin lines rising above the background in Figs.~\ref{fig:u-evol-c}-\ref{fig:u-evol-d}. The evolution is not always monotonic, since there are grains
 that first increase their mass and then tend to zero (lines that bend toward the $z=0$ plane in the plots). In addition,
 a small number of grains undergo different growth phases marked by sudden slope changes.

\begin{figure}[ht]
\begin{center}
\subfigure[$u=0.05$]{\includegraphics[width=0.45\linewidth]{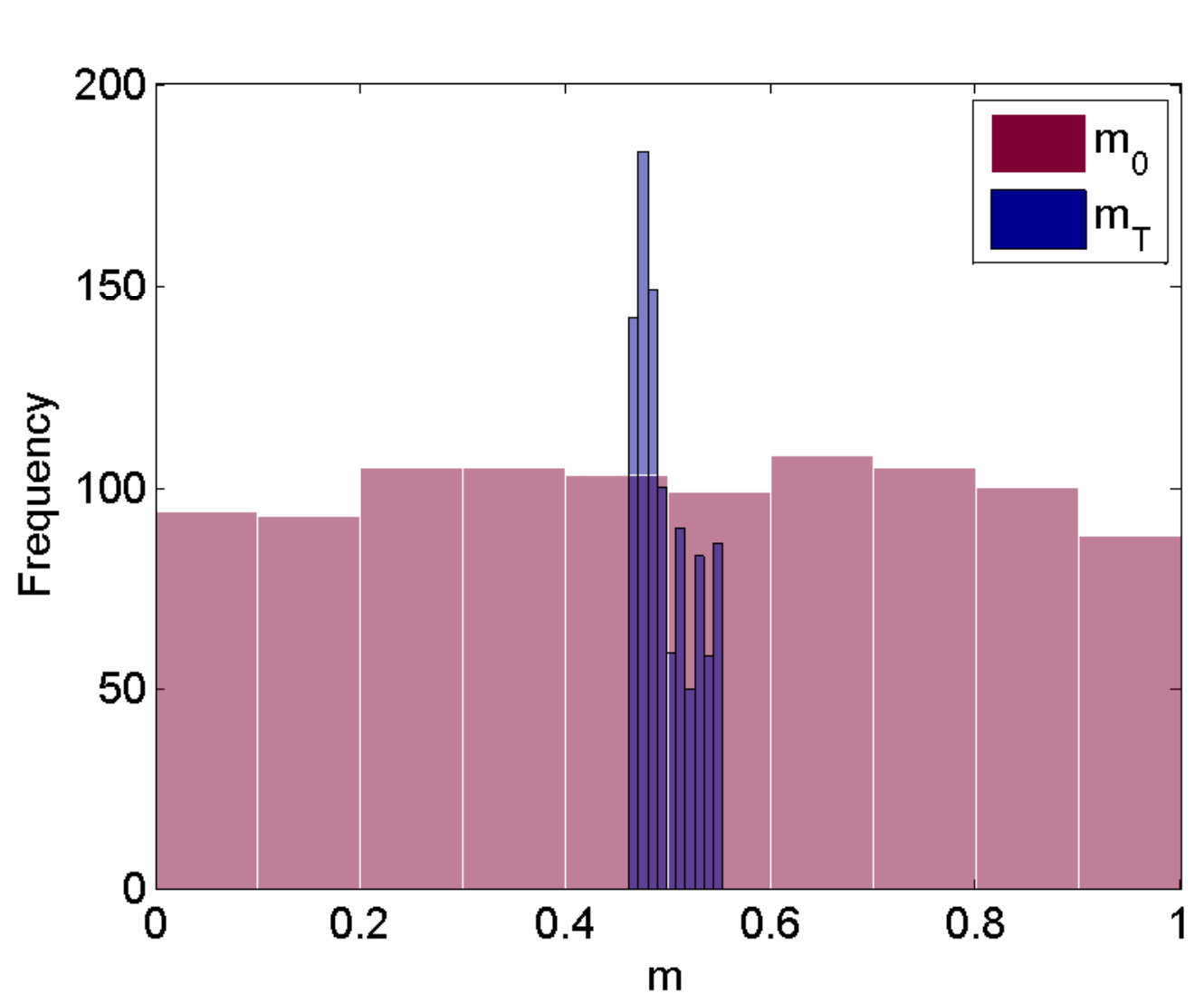}\label{fig:histo-a}}
\subfigure[$u=1$]{\includegraphics[width=0.45\linewidth]{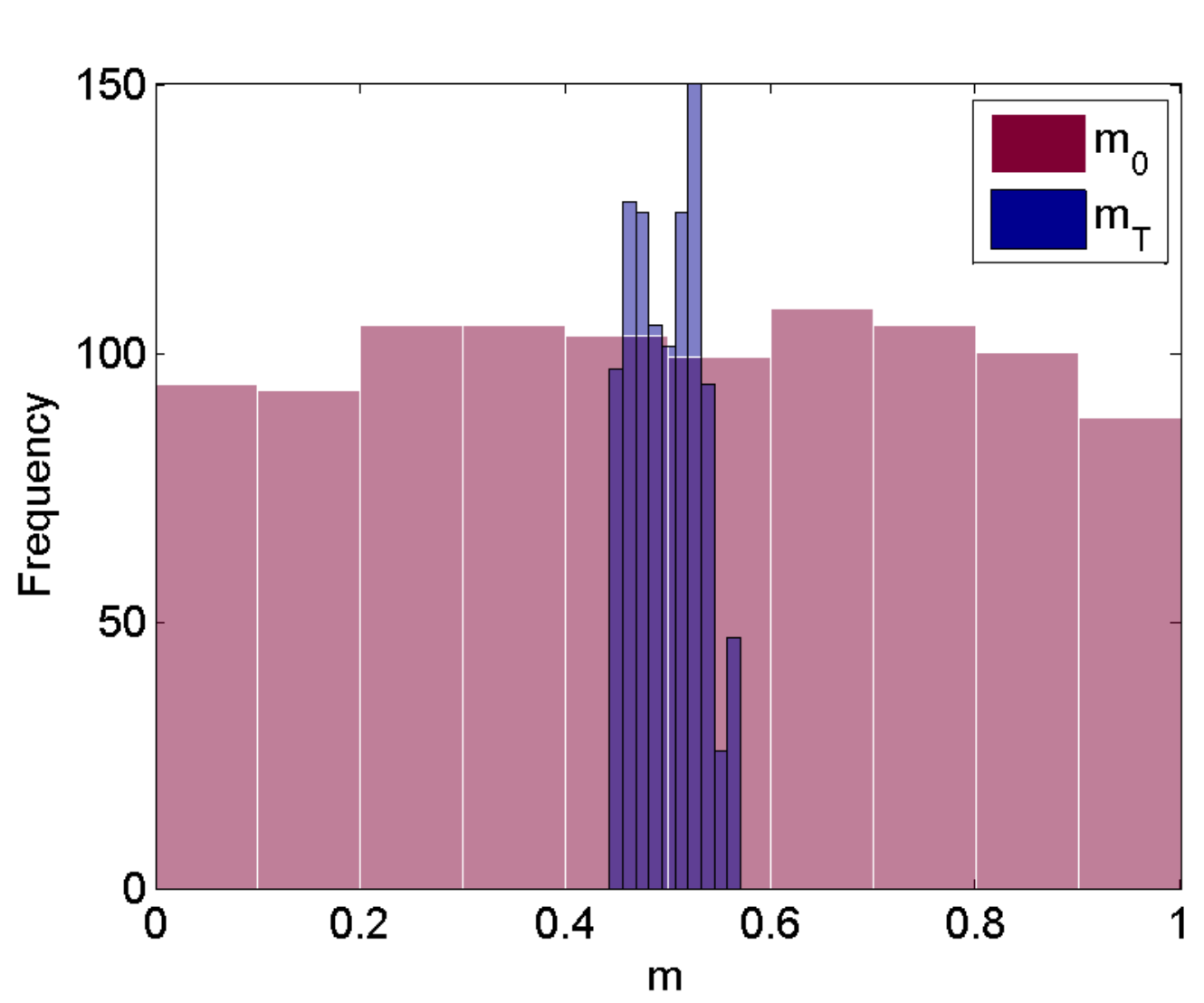}\label{fig:histo-b}}
\subfigure[$u = 1.5$]{\includegraphics[width=0.45\linewidth]{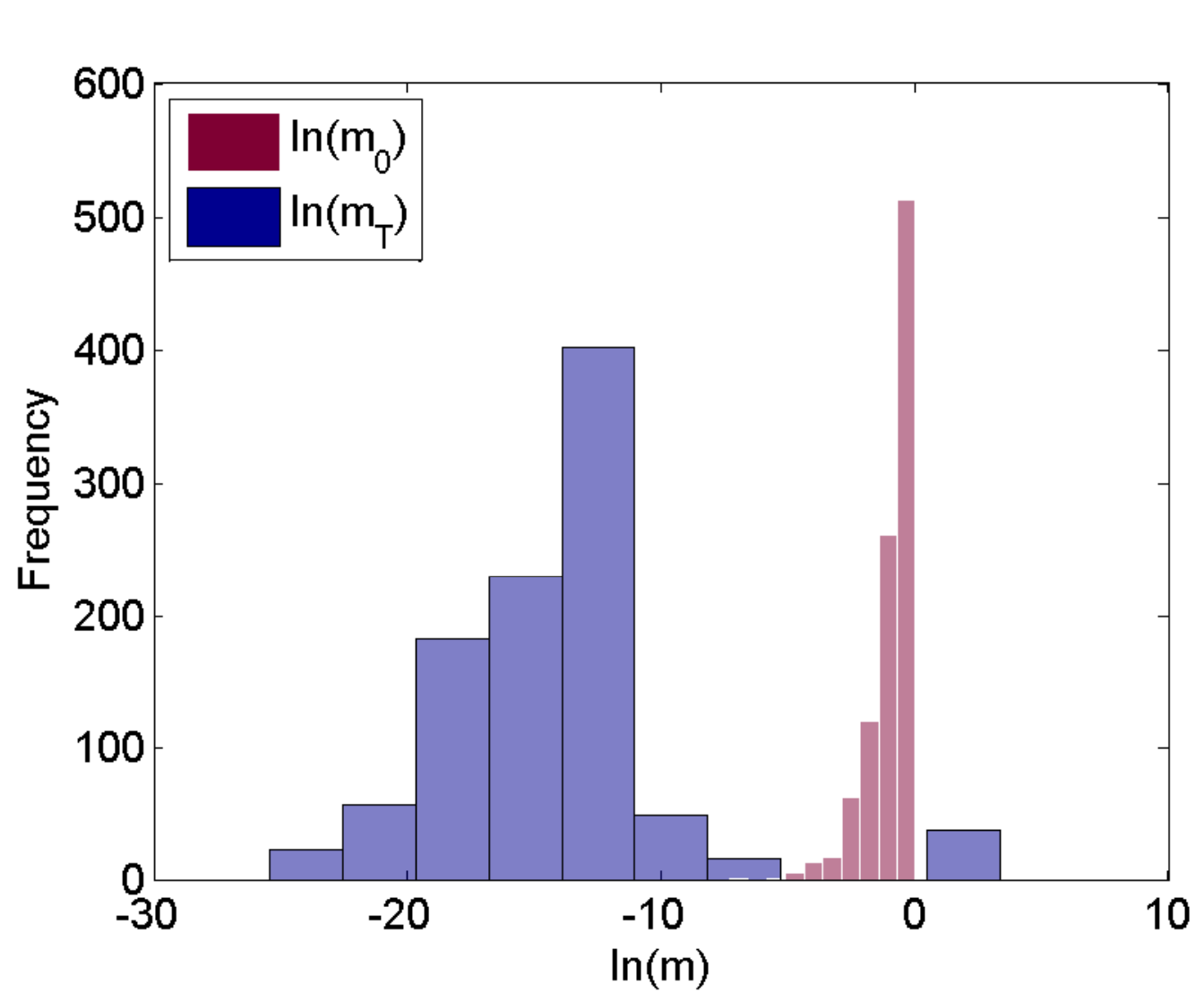}\label{fig:histo-c}}
\subfigure[$u =2$]{\includegraphics[width=0.45\linewidth]{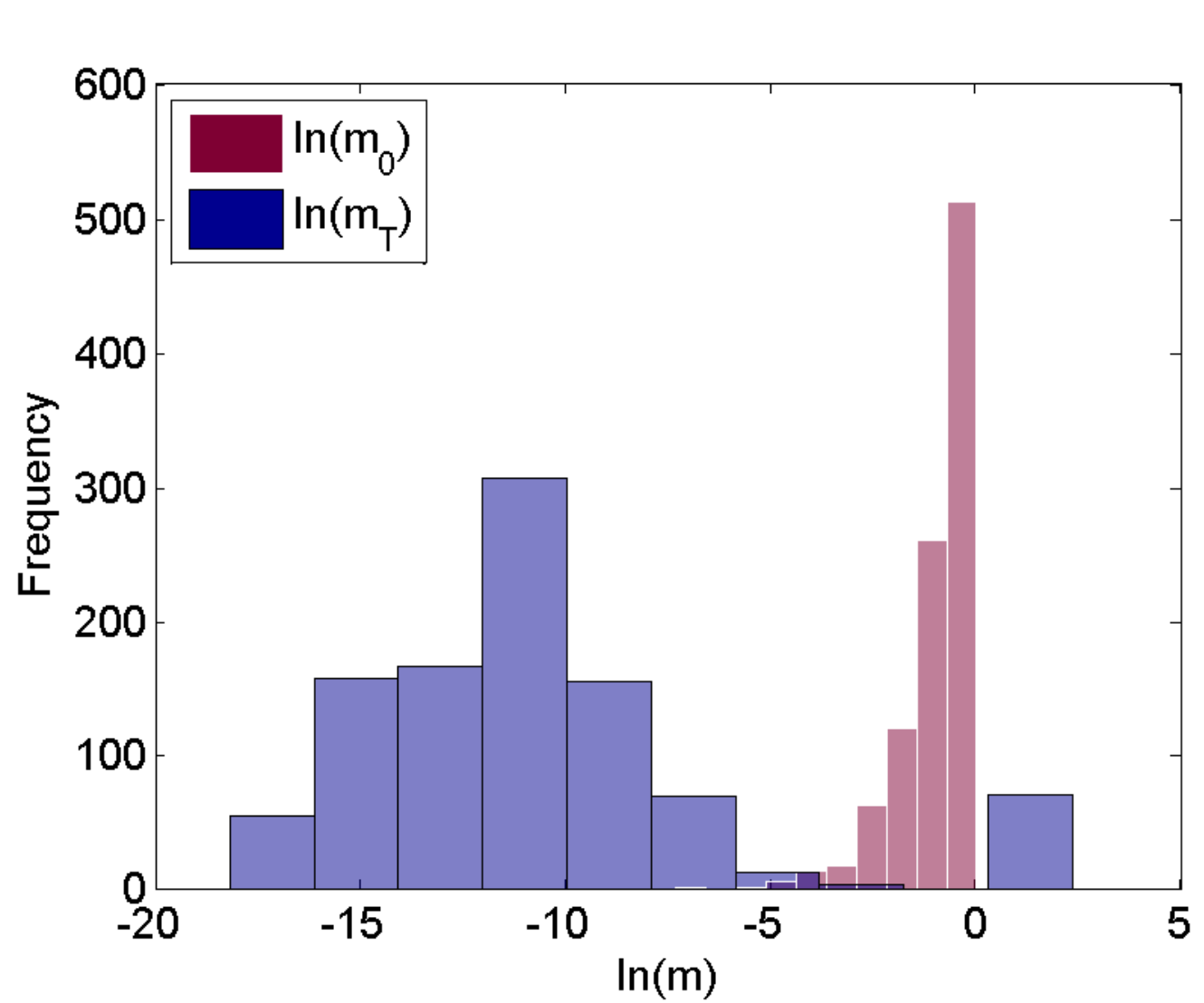}\label{fig:histo-d}}
\end{center}
 \caption{Histograms of the initial mass distribution (bars with white edges, red bar-face color online) and final mass distribution
 (bars with dark edges, blue bar-face color online)
 of a system comprising $N=10^3$ grain masses for different values of $u$ based on the solution of
 ODE system~\eqref{eq:mass-transN}.  The final time is $T=10^3$ for $u \le 1$ and $T=10^4$ for $u>1$.
 The initial mass configuration is drawn from the uniform distribution $U(0,1)$.  For $u>1$ the histogram of the grain mass logarithm
 is shown to improve resolution. }
  \label{fig:u-histo}
\end{figure}

 The patterns  described above that pertain to the equilibrium distribution
 are confirmed by comparing the histograms of the initial and final mass distributions shown in Fig.~\ref{fig:u-histo}.
 The same qualitative patterns are obtained if we use a lognormal initial mass distribution (not shown).
 In the lognormal case,
 the growth-decay regime is established at $u=1$.  This is due to the fact that the
 equilibrium is determined by both $u$ and the
 magnitude of the grain masses. The lognormal distribution is broader than the $U(0,1)$
  distribution, and thus includes some higher  mass
values which drive the system into the growth-decay regime at lower $u$.



 If $u$ is kept in the Arrhenius factor the system involves a redundant degree of freedom as discussed above.
 We can use $u$, however, as a control parameter  to investigate
 the impact of temperature on the evolution of the system.
 We can then normalize the grain masses, e.g., by their
 mean value, to reduce the number of parameters by one.
As a special case, we can   set $u \to 1/ m_{\rm tot}$ and replace $u \, m_{i}(t)$ with $m_{i}(t)$.
With this normalization, it holds that $\sum_{i=1}^{N} m_{i}(t) =1$ for all $t$.
Thus, the masses can be viewed
as probabilities for different states, and eq.~\eqref{eq:mass-transN} becomes a \emph{master equation}.

\section{Conclusions}
\label{sec:conclu}
We investigated a nonlinear system of ordinary differential equations that describes mass exchange
 between grains.
The exchange is governed by Arrhenius-type transition rates with an activation energy
that depends linearly on grain mass.
We identified the equilibrium states of a two-grain system which are defined by the
linear equipartition nullcline and the curvilinear mass separation nullcline.  The system exhibits
diffusive and growth-decay regimes. In the diffusive regime, the equilibrium state of the system is
 equipartition unless the diffusive process is arrested at the separation nullcline; in this case the
 mass transfer stops before the two masses are equalized. In the growth-decay regime, the larger grain grows
 at the expense of the smaller grain that shrinks; the mass transfer is again stopped at the separation
 nullcline.
We derived linearized approximations which are
valid in parts of the diffusive (``high temperature'') and growth-decay (``low temperature'') regimes respectively.
The linear approximations
provide a qualitative understanding of the system which is more accurate in the diffusive regime.
The linear approximations, however, miss the trapping by the separation nullcline.
We also constructed the equilibrium phase diagram of the model based on the long-time mass
difference between the grains obtained by numerical
solutions of the system~\eqref{eq:two-grain-nou}.

 Numerical solutions of a two-grain system with additive white noise in the mass transfer rates
 reveal that the
 equilibrium states fluctuate around the deterministic equilibrium points.
 The noise has significant impact if the initial conditions are
 close to the equipartition nullcline and in the growth-decay regime.
 In this case the noise can reverse the direction of mass growth from the bigger to the smaller grain.
 We also analyzed an $N$-grain system with periodic boundary conditions using numerical solutions
 based on the fourth-order Runge-Kutta method. We use a variable $u$ and initial masses drawn from the
 $U(0,1)$ distribution.
 We established that diffusive and growth-decay regimes also exist; the former are obtained for lower
 values of $u$ and the latter for higher values. These regimes exhibit richer patterns of
 grain mass evolution than the respective  two-grain-system regimes and
 will be further investigated in further research.

Based on the analysis above, we suggest that the growth-decay
 regime is related to abnormal grain growth
that is observed  in sintering and leads to grain coarsening.
The behavior of many-grain systems as well as grain coalescence,
which are important for sintering applications,
 are being investigated in ongoing research by our group.
Finally, our model could be useful as a component
of kinetic models of wealth exchange~\cite{Bouchaud00,During08}.
In this context, the activation parameter is proportional to the average wealth of the
agents (grains) and to increased ability for exchanges (lower $u$), whereas mass conservation corresponds to total wealth conservation~\cite{Dragu00}.
The existence of two distinct regimes, one corresponding to diffusion of wealth
among the agents and
the other to wealth accumulation by few agents, is an intriguing feature of the model.

\section*{Acknowledgements}
This work has been funded by the project \emph{NAMCO: Development of High Performance Alumina Matrix Nanostructured
Composites}. NAMCO is implemented under the
``THALIS'' Action of the operational programme ``Education
and Lifelong Learning'' and is co-funded by the European
Social Fund and National Resources. We also acknowledge the contributions of former
students Spyros Blanas and Ioannis Kardaras in earlier numerical investigations.

\section*{References}

\bibliographystyle{elsarticle-num-names}

\begin{thebibliography}{23}
\providecommand{\natexlab}[1]{#1}
\providecommand{\url}[1]{\texttt{#1}}
\providecommand{\urlprefix}{URL }
\expandafter\ifx\csname urlstyle\endcsname\relax
  \providecommand{\doi}[1]{doi:\discretionary{}{}{}#1}\else
  \providecommand{\doi}[1]{doi:\discretionary{}{}{}\begingroup
  \urlstyle{rm}\url{#1}\endgroup}\fi
\providecommand{\bibinfo}[2]{#2}

\bibitem[{Cetinel et~al.(2013)Cetinel, Kayacan, and Ozaydin}]{Cetinel13}
\bibinfo{author}{H.~Cetinel}, \bibinfo{author}{O.~Kayacan},
  \bibinfo{author}{D.~Ozaydin},  \bibinfo{journal}{Physica A} \bibinfo{volume}{392}
  (\bibinfo{year}{2013}) \bibinfo{pages}{4121}.

\bibitem[{Rub{\'{\i}} and Gadomski(2003)}]{Rubi03}
\bibinfo{author}{J.~M. Rub{\'{\i}}}, \bibinfo{author}{A.~Gadomski},
  \bibinfo{journal}{Physica A}
  \bibinfo{volume}{326} (\bibinfo{year}{2003})
  \bibinfo{pages}{333}.

\bibitem[{Avrami(1939)}]{avrami39}
\bibinfo{author}{M.~Avrami},  \bibinfo{journal}{J. Chem. Phys.}
  \bibinfo{volume}{7} (\bibinfo{year}{1939}) \bibinfo{pages}{1103}.

\bibitem[{Avrami(1941)}]{avrami41}
\bibinfo{author}{M.~Avrami},  \bibinfo{journal}{J. Chem. Phys.} \bibinfo{volume}{9} (\bibinfo{year}{1941})
  \bibinfo{pages}{177}.

\bibitem[{Eggers(1998)}]{Eggers98}
\bibinfo{author}{J.~Eggers},  \bibinfo{journal}{Phys. Rev. Lett.}
  \bibinfo{volume}{80} (\bibinfo{year}{1998})
  \bibinfo{pages}{2634}.

\bibitem[{J.H.~Yao and Grant(2003)}]{Yao93}
\bibinfo{author}{H.~G. J.H.~Yao, K. R.~Elder}, \bibinfo{author}{M.~Grant},
  \bibinfo{journal}{Phys. Rev. B} \bibinfo{volume}{47}
  (\bibinfo{year}{2003}) \bibinfo{pages}{14110}.

\bibitem[{Pototsky et~al.(2014)Pototsky, Thiele, and Archer}]{Pototsky14}
\bibinfo{author}{A.~Pototsky}, \bibinfo{author}{U.~Thiele},
  \bibinfo{author}{A.~J. Archer},  \bibinfo{journal}{Phy. Rev. E}
  \bibinfo{volume}{89} (\bibinfo{year}{2014})
  \bibinfo{pages}{032144}.

\bibitem[{Maximenko et~al.(2012)Maximenko, Kuzmov, Grigoryev, and
  Olevsky}]{Max12}
\bibinfo{author}{A.~Maximenko}, \bibinfo{author}{A.~Kuzmov},
  \bibinfo{author}{E.~Grigoryev}, \bibinfo{author}{E.~Olevsky},
  \bibinfo{journal}{J. Amer. Cer. Soc.}
  \bibinfo{volume}{95} (\bibinfo{year}{2012})
  \bibinfo{pages}{2383}.

\bibitem[{Martin et~al.(2006)Martin, Schneider, Olmos, and Bouvard}]{martin06}
\bibinfo{author}{C.~L. Martin}, \bibinfo{author}{L.~C.~R. Schneider},
  \bibinfo{author}{L.~Olmos}, \bibinfo{author}{D.~Bouvard},
  \bibinfo{journal}{Scripta Mater.} \bibinfo{volume}{55}
  (\bibinfo{year}{2006}) \bibinfo{pages}{425}.

\bibitem[{Olmos et~al.(2009)Olmos, Martin, Bouvard, Bellet, and
  Michielz}]{martin09}
\bibinfo{author}{L.~Olmos}, \bibinfo{author}{C.~L. Martin},
  \bibinfo{author}{D.~Bouvard}, \bibinfo{author}{D.~Bellet},
  \bibinfo{author}{M.~D. Michielz},  \bibinfo{journal}{J. Amer.
  Cer. Soc.} \bibinfo{volume}{92}
  (\bibinfo{year}{2009}) \bibinfo{pages}{1492}.



\bibitem[{{Cahn} and {Hilliard}(1958)}]{CH58}
\bibinfo{author}{J.~W. {Cahn}}, \bibinfo{author}{J.~E. {Hilliard}},
   \bibinfo{journal}{J. Chem. Phys.} \bibinfo{volume}{28}
  (\bibinfo{year}{1958}) \bibinfo{pages}{258}.

\bibitem[{Langer(1986)}]{Langer86}
\bibinfo{author}{J.~Langer}, \bibinfo{title}{Models of pattern formation in
  first–order phase transitions}, in: \bibinfo{editor}{G.~Grinstein},
  \bibinfo{editor}{G.~Mazenko} (Eds.), \bibinfo{booktitle}{Directions in
  Condensed Matter Physics}, \bibinfo{publisher}{World Scientific},
  \bibinfo{address}{Philadelphia}, \bibinfo{pages}{165–--186},
  \bibinfo{year}{1986}.

\bibitem[{Nastar(2014)}]{Nastar14}
\bibinfo{author}{M.~Nastar}, \bibinfo{journal}{Phys. Rev. B}
  \bibinfo{volume}{90} (\bibinfo{year}{2014})
  \bibinfo{pages}{144101}.

\bibitem[{Sopicka-Lizer et~al.(2013)Sopicka-Lizer, Duran, Gocmez, Pawlik,
  Mikuskiewicz, and MacKenzie}]{Sopicka13}
\bibinfo{author}{M.~Sopicka-Lizer}, \bibinfo{author}{C.~Duran},
  \bibinfo{author}{H.~Gocmez}, \bibinfo{author}{T.~Pawlik},
  \bibinfo{author}{M.~Mikuskiewicz}, \bibinfo{author}{K.~MacKenzie},
  \bibinfo{journal}{Ceram. Intern.}
  \bibinfo{volume}{39} (\bibinfo{year}{2013})
  \bibinfo{pages}{4269}.

\bibitem[{Hillert(1965)}]{Hillert65}
\bibinfo{author}{M.~Hillert},  \bibinfo{journal}{Acta Metallur.}
  \bibinfo{volume}{13} (\bibinfo{year}{1965})
  \bibinfo{pages}{227}.

\bibitem[{Hristopulos et~al.(2006)Hristopulos, Leonidakis, and
  Tsetsekou}]{dth06}
\bibinfo{author}{D.~T. Hristopulos}, \bibinfo{author}{L.~Leonidakis},
  \bibinfo{author}{A.~Tsetsekou},  \bibinfo{journal}{Eur. Phys. J. B}
  \bibinfo{volume}{50} (\bibinfo{year}{2006})
  \bibinfo{pages}{83}.


\bibitem[{King et~al.(2003)King, Billingham, Otto}]{king03}
\bibinfo{author}{A.C. King} , \bibinfo{author}{J. Billingham},
\bibinfo{author}{S.R. Otto}, \bibinfo{title}{Differential Equations},
\bibinfo{publisher}{Cambridge},
  \bibinfo{address}{Cambridge}, \bibinfo{year}{2003}.

\bibitem[{Press et~al.(1997)}]{Press97}
\bibinfo{author}{W.~H. Press}, et~al., \bibinfo{title}{Numerical Recipes in
  Fortran 77, Volume 1}, \bibinfo{publisher}{Cambridge},
  \bibinfo{address}{Cambridge}, \bibinfo{year}{1997}.

\bibitem[{Zwanzig(2001)}]{Zwanzig01}
\bibinfo{author}{R.~W. Zwanzig}, \bibinfo{title}{Nonequilibrium Statistical
  Mechanics}, \bibinfo{publisher}{Oxford Univ. Press}, \bibinfo{year}{2001}.

\bibitem[{Uhlenbeck and Ornstein(1930)}]{UO30}
\bibinfo{author}{G.~E. Uhlenbeck}, \bibinfo{author}{L.~S. Ornstein},
  \bibinfo{journal}{Phys. Rev.} \bibinfo{volume}{36} (\bibinfo{year}{1930})
  \bibinfo{pages}{823}.

\bibitem[{Gillespie(1996)}]{Gillespie96}
\bibinfo{author}{D.~T. Gillespie},
 \bibinfo{journal}{Phys. Rev. E} \bibinfo{volume}{54} (\bibinfo{year}{1996})
  \bibinfo{pages}{2084}.

\bibitem[{Bouchaud and M{\'e}zard(2000)}]{Bouchaud00}
\bibinfo{author}{J.-P. Bouchaud}, \bibinfo{author}{M.~M{\'e}zard},
  \bibinfo{journal}{Physica A}
  \bibinfo{volume}{282} (\bibinfo{year}{2000})
  \bibinfo{pages}{536}.

\bibitem[{D\"uring et~al.(2008)D\"uring, Matthes, and Toscani}]{During08}
\bibinfo{author}{B.~D\"uring}, \bibinfo{author}{D.~Matthes},
  \bibinfo{author}{G.~Toscani},
  \bibinfo{journal}{Phys. Rev. E} \bibinfo{volume}{78} (\bibinfo{year}{2008}) \bibinfo{pages}{056103}.

\bibitem[{Dragulescu and Yakovenko(2000)}]{Dragu00}
\bibinfo{author}{A.~Dragulescu}, \bibinfo{author}{V.~M. Yakovenko},
\bibinfo{journal}{Eur. Phys. J. B} \bibinfo{volume}{17}
  (\bibinfo{year}{2000}) \bibinfo{pages}{723}.

\end{thebibliography}

\end{document}